\documentclass[twocolumn,pra,superscriptaddress,showkeys,longbibliography, colorlinks=true]{revtex4-2}
\usepackage{bm,mathtools,xcolor,booktabs}
\usepackage[colorlinks, linkcolor=blue, citecolor=blue, breaklinks=true]{hyperref}
\usepackage[nameinlink,capitalize]{cleveref}
\DeclareMathAlphabet{\mathcal}{OMS}{cmsy}{m}{n}

\begin{document}
\title{Saturable nonlinearity induced quantum correlations in optomechanics}

\author{D.R. \surname{Kenigoule Massembele}}
\email{kenigoule.didier@gmail.com}
\affiliation{Department of Physics, Faculty of Science,
University of Ngaoundere, P.O. Box 454, Ngaoundere, Cameroon}

\author{E. \surname{Kongkui Berinyuy}}
\email{emale.kongkui@facsciences-uy1.cm}
\affiliation{Department of Physics, Faculty of Science, University of Yaounde I, P.O.Box 812, Yaounde, Cameroon}

\author{P. Djorwé}
\email{djorwepp@gmail.com}
\affiliation{Department of Physics, Faculty of Science,
University of Ngaoundere, P.O. Box 454, Ngaoundere, Cameroon}
\affiliation{Stellenbosch Institute for Advanced Study (STIAS), Wallenberg Research Centre at Stellenbosch University, Stellenbosch 7600, South Africa}

\author{A.-H. Abdel-Aty}
\affiliation{Department of Physics, College of Sciences, University of Bisha, Bisha 61922, Saudi Arabia}
\affiliation{Physics Department, Faculty of Science, Al-Azhar University, Assiut 71524, Egypt}

\author{M.R. Eid}
\email{mohamed.eid@nbu.edu.sa}
\affiliation{Center for scientific research and entrepreneurship, Northern border University, Arar 73213, Saudi Arabia}

\author{R. Altuijri}
\email{raaltuwagry@pnu.edu.sa}
\affiliation{Department of Physics, College of Science, Princess Nourah bint Abdulrahman University, P. O. Box 84428, Riyadh 11671, Saudi Arabia}

\author{S. G. \surname{Nana Engo}}
\email{serge.nana-engo@facsciences-uy1.cm}
\affiliation{Department of Physics, Faculty of Science, University of Yaounde I, P.O.Box 812, Yaounde, Cameroon}

\begin{abstract}
We propose a scheme that induces quantum correlations in optomtomechanical systems. Our benchmark system consists of two optically coupled optical cavities which interact with a common mechanical resonator. The optical cavities host saturable nonlinearity which triggers either gain or losses in each cavity. Without these nonlinearities, there are no quantum correlations, i.e., entanglement and steering, generated in the system. By turning on the nonlinearities, gain and losses are switched on, enabling flexible generation of both  quantum entanglement and quantum steering in our proposal. These generated quantum correlations seem to be insensitive to the induced gain, while the induced losses through saturation effect efficiently enhance quantum correlations. Moreover, the robustness of the generated quantum correlations against thermal fluctuations is further improved under nonlinear saturation scenario. This work suggests a way of using nonlinear saturation effects to engineer quantum correlations even at room temperature, which are useful for quantum information processing, quantum computational tasks, and quantum technologies.
\end{abstract}

\pacs{ 42.50.Wk, 42.50.Lc, 05.45.Xt, 05.45.Gg}
\keywords{Quantum correlations, optomechanics, saturable nonlinearity.}

\maketitle
\date{\today}

\section{Introduction}\label{intro}

Cavity optomechanical systems (COMs), which couple electromagnetic fields to mechanical objects, are versatile platforms that bridge classical and quantum regimes, offering profound insights into the interplay between light and mechanical vibrations. These systems have fostered advances across fundamental quantum sciences and emerging related technologies. In the classical regime, COMs underpin applications such as chaos generation \cite{Monifi2016,Navarro2017,Mbok2024}, synchronization \cite{Sheng2020,Djor2020,Rodrigues2021}, and sensors sensitivity \cite{Djorwe2019,Li2021,Yan2023,Dj2024}. From quantum  perspectives, several achievements have been put forward  including ground-state cooling \cite{Brubaker2022,Wang2024,Cao2025}, nonclassical state generation \cite{Wise2024}, and the generation of quantum correlations \cite{Bemani2019,Rostand2025,Emale.2025}, just to name few. 

The quest of robust quantum correlations, i.e., entanglement and steering, has emerged as conerstone of modern quantum technologies owing to their intersting roles in quantum information processing \cite{Slussarenko2019}, metrology \cite{Pezz2018}, and advanced sensing \cite{Xia2023}. While early approaches leveraged intrinsic optomechanical interactions  \cite{Vitali2007,Pater2007} or parametric modulation \cite{Farace2012,Mari2012}, recent strategies increasingly make use of nonlinear effects \cite{Agasti2024,Rostand2024}, dark mode engineering \cite{Lai2022}, and synthetic gauge fields \cite{Zhai2023}. Despite these significant progresses made on generating quantum correlations based on the aforementioned approaches, engineering abundant quantum correlations having high robustness against thermal noise is still challenging. Recent breakthroughs in nonlinear systems have demonstrated that saturable nonlinearity can effectively suppress noise amplification near exceptional points (EPs) \cite{Hassan2015,Bai2022,Gu2024}, owing to a restoration of the eigenstate orthogonality leading to a finite Petermann factors. This motivates our investigation of saturable nonlinearity as a mechanism to enhance quantum correlations in optomechanical systems.

Our proposed benchmark system is a double-cavity optomechanical system, which are optically coupled and they interact with a common mechanical resonator. Each cavity host saturable nonlinearity that induces either gain or losses in a given cavity. The key findings of our  work reveal that: (i) bipartite entanglement and two-way steering are triggered in the system through saturable nonlinearity, (ii) the saturated loss is mainly responsible for the enhancement of the quantum correlations, and (iii) the generated quantum correlations are more robust to thermal noise in the presence of the saturated loss, enabling room-temperature quantum correlations. Our findings suggest insights into the role of saturable nonlinearities in the generation of quantum correlations, which are interesting resources for quantum information processing, quantum communication protocols,  and quantum computational tasks.

The rest of the paper is organized as follows: \Cref{sec:model} describes our proposed scheme and the related dynamical equations. \Cref{sec:entangl} analyzes the features of saturable nonlinearities induced quantum correlations through entanglement and steering metrics. \Cref{sec:concl} summarizes our key findings and outlines future research directions together with some implications for quantum technologies.

\section{Model and dynamics} \label{sec:model}

\begin{figure}[htb]
  \centering
  \includegraphics[width=0.8\linewidth]{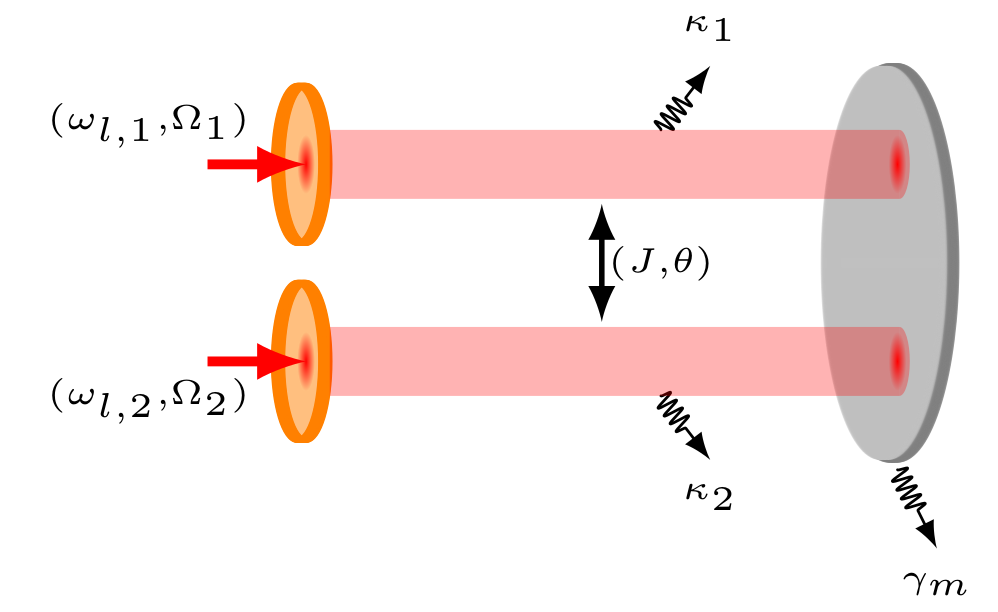}
  \caption{Sketch of the double-cavity optomechanical system. Two optical modes (frequency $\omega_{c,j}$, decay rate $\kappa_j$) are coupled via photon hopping ($J$) and interact with a common mechanical resonator (frequency $\omega_m$, damping rate $\gamma_m$). Saturable gain ($g_s$) and loss ($f_s$) are engineered in the respective cavities.}
  \label{fig:Fig1}
\end{figure}

Our benchmark system is an optomechanical structure consisting of two optical cavities interacting with a common mechanical resonator through optomechanical coupling. The two cavities are optically coupled through a photon hopping rate $J$, and each of them hosts either a gain or a loss induced by the saturable nonlinearity related to the driving field.  Similar configuration, illustrated in \Cref{fig:Fig1}, has been recently used to generate two-colour interferometry and switching in optomechanis via dark mode excitation \cite{Lake2020}.

The dynamical state of our system is captured by the Hamiltonian ($\hbar=1$),
\begin{equation}\label{eq:ham1}
 H =H_{\rm{OM}}+H_{\rm{int}}+H_{\rm{drive}}+H_{\rm{sat}}, 
 \end{equation}
where
\begin{eqnarray}  \label{eq:ham2}
H_{\rm{OM}}&:=& \omega_m b^\dagger b + \sum_{j=1,2} \Delta_{c,j} a_j^\dagger a_j +  g_j(b^\dagger + b)a_j^\dagger a_j, \\
H_{\rm{int}}&:=&J(a_1^\dagger a_2+ a_1 a_2^\dagger ) , \\
H_{\rm{drive}}&:=&\sum_{j=1,2} \Omega_j(a_j^\dagger + a_j), \\
H_{\rm{sat}}&:=&ig_s a_1^\dagger a_1 - i f_s a_2^\dagger a_2.
\end{eqnarray} 
The terms $H_{\rm{OM}}$, $H_{\rm{int}}$, $H_{\rm{drive}}$ and $H_{\rm{sat}}$ represent the optomechanical Hamiltonian, the optical interaction between the cavities, the driving fields, and the saturation gain/loss effects within the system, respectively. The operators $a_j$ ($a_j^\dagger$) and $b$ ($b^\dagger$) are the annihilation (creation) operators of the $j^{th}$ optical mode  and  mechanical mode. The $g_j$ stands for the optomechanical coupling between the mechanical resonator and the $j^{th}$ optical mode. The $j^{th}$ driving field (of frequency $\omega_{l,j}$) is described by the amplitude $\Omega_j$, and the frequency detuning is defined by $\Delta_{c,j}=\omega_{c,j}-\omega_{l,j}$. The parameters $g_s$ and $f_s$ denote the saturable gain and loss respectively, and are expressed as follows,
\begin{equation}\label{eq:sat}
g_s=\frac{g_0}{1+|\alpha_1|^2}, \hspace*{1cm} f_s=\frac{f_0}{1+|\alpha_2|^2},
\end{equation}
where $g_0$ and $f_0$ are the small-signal gain and loss coefficients, and $\alpha_j$ is the steady-state amplitude of the $j$-th optical mode.

\subsection{Quantum Langevin Equations}

By taking into account the dissipations, the dynamical state of our system is captured by the following Quantum Langevin Equations (QLEs),
\begin{eqnarray}\label{eq:qle}
\dot{a}_1&=-\left(i\left(\Delta_{c,1} +g_1(b^\dagger+b)\right) - g \right)a_1-iJ a_2 -i\Omega_1 \nonumber \\&+\sqrt{2\kappa_1}a_1^{in}, \\
\dot{a}_2&=-\left(i\left(\Delta_{c,2} +g_2(b^\dagger+b)\right) + f \right)a_2-iJ a_1 -i\Omega_2 \nonumber \\&+\sqrt{2\kappa_2}a_2^{in}, \\
\dot{b}&=-\left(i\omega_m + \gamma_m \right)b - i\sum_{j=1,2} g_j a_j^\dagger a_j +\sqrt{2\gamma_m}b^{in}, 
\end{eqnarray} 
where $g=g_s - \kappa_1$ and $f=f_s + \kappa_2$ represent the net modal gain and loss in the first and second cavities, respectively. The quantum noise operators $a_j^{in}$ and $b^{in}$ have zero mean and are characterized by the following autocorrelation functions,
\begin{eqnarray}\label{eq:noise}
&\langle a_j^{in}(t)a_j^{in\dagger}(t') \rangle &=\delta(t-t'), \hspace*{0.1cm} \langle a_j^{in\dagger}(t)a_j^{in}(t') \rangle = 0, \\
&\langle b^{in}(t)b^{in\dagger}(t') \rangle &=(n+1)\delta(t-t'), \\ 
&\langle b^{in\dagger}(t)b^{in}(t') \rangle &= n \delta(t-t').
\end{eqnarray}
In these autocorrelation functions, $n$ represents the equilibrium phonon occupation number of the mechanical resonator, and is defined as $n=[\rm exp(\frac{\hbar \omega_m}{k_bT})-1]^{-1}$, where $\rm k_b$ is the Boltzmann constant and $T$ the temperature. 

\subsection{Linearization and fluctuation dynamics}

Due to the presence of nonlinearities, direct analytical treatment of the full QLEs is challenging. However, under the assumption of weak driving fields (such that $g_s \approx g_0$ and $f_s \approx f_0$), we linearize the equations by expressing each operator as a sum of its steady-state mean value and a small quantum fluctuation. This leads to the following steady-state equations,
\begin{eqnarray}\label{eq:means}
\dot{\alpha}_1&=-\left(i{\Delta}_1 - g \right)\alpha_1-iJ \alpha_2 - i\Omega_1, \\
\dot{\alpha}_2&=-\left(i{\Delta}_2 + f \right)\alpha_2-iJ  \alpha_1 - i\Omega_2, \\
\dot{\beta}&=-\left(i\omega_m + \gamma_m \right)\beta - i\sum_{j=1,2} g_j |\alpha_j|^2,
\end{eqnarray}
and a fluctuations dynamics,
\begin{eqnarray}\label{eq:fluct}
\delta \dot{a}_1&=-\left(i{\Delta}_1 - g \right)\delta a_1 - iG_1(\delta b^\dagger + \delta b)-iJ \delta a_2  \nonumber \\&+\sqrt{2\kappa_1}a_1^{in}, \\
\delta \dot{a}_2&=-\left(i{\Delta}_2 + f \right)\delta {a}_2 - iG_2(\delta b^\dagger + \delta b) -iJ  \delta a_1  \nonumber \\&+\sqrt{2\kappa_2}a_2^{in},\\
\delta \dot{b}&=-\left(i\omega_m + \gamma_m \right)\delta {b} - i\sum_{j=1,2} (G_j \delta a_j^\dagger + G_j^* \delta a_j) \nonumber \\&+\sqrt{2\gamma_m}b^{in},
\end{eqnarray}
where ${\Delta}_j=\Delta_{c,j} +g_j(\beta^\ast +\beta)$ and $G_{j}=g_j \alpha_j$ are the effective detuning and optomechanical coupling.

\subsection{Quadrature representation and covariance matrix}

In order to quantify quantum correlations,  we define the following position and momentum quadrature operators: $\delta X_{o} =\frac{\delta o^\dagger + \delta o }{\sqrt{2}}$, $\delta Y_{o} =i\frac{\delta o^\dagger - \delta o}{\sqrt{2}}$, with $o\equiv a_j,b$ with the related noise quadratures $\delta X_{o}^{in} =\frac{\delta o^{\dagger in} + \delta o^{in}}{\sqrt{2}}$, $\delta Y_{o}^{in} =i\frac{\delta o^{\dagger in} - \delta o^{in}}{\sqrt{2}}$. These quadratures can be used to rewrite the above fluctuation dynamical equations in the following compact form,
\begin{equation}\label{eq:quadra}
\dot{\mathbf{X}}=\mathbf{X} \mathbf{M}+\sqrt{2\mu_j} \mathbf{z}^{in},
\end{equation}
where $\mathbf{X} = (\delta X_{a_1}, \delta Y_{a_1}, \delta X_{a_2}, \delta Y_{a_2}, \delta X_b, \delta Y_b)^\top$ is the vector of quadrature fluctuations,  $\mu_j\equiv \kappa_j, \gamma_m$, and $\mathbf{z}^{in}=( X_{a_1}^{in}, Y_{a_1}^{in}, X_{a_2}^{in}, Y_{a_2}^{in}, X_{b}^{in}, Y_{b}^{in},)^{\top}$ collects the corresponding noise terms. The drift matrix $\mathbf{M}$ is given by,
\begin{equation}\label{eq:matrix}
\mathbf{M}=\begin{pmatrix}
g&\Delta_1&0&J&0&0 \\
-\Delta_1&g&-J&0&-2G_{1}&0 \\
0&J&-f&\Delta_2&0&0 \\
-J&0&-\Delta_2&-f&-2G_{2}&0 \\
0&0&0&0&-\gamma_m&\omega_m \\
-2G_{1}&0&-2G_{2}&0&-\omega_m&-\gamma_m
\end{pmatrix},
\end{equation}
with ($^{\top}$) being the transpose symbol. In our investigation, we will assume for simplicity that $G_j$ is real.

Quantum correlations are quantified from the covariance matrix $V$, which obeys the following Lyapunov differential equation, 
\begin{equation}\label{eq:dlyap}
\dot{\mathbf{V}} = \mathbf{M}\mathbf{V} + \mathbf{V}\mathbf{M}^\top + \mathbf{D}.
\end{equation}
The elements of this covariance matrix can also be defined as $V_{ij}=\frac{\langle u_i u_j  + u_j u_i \rangle}{2}$, and the diffusion matrix associated with noise correlation functions is $D=diag[{\kappa_1},  {\kappa_1}, {\kappa_2}, {\kappa_2}, {\gamma_m}(2n + 1),{\gamma_m}(2n + 1)]$.  The stability of the system is ensured by requiring all eigenvalues of $\mathbf{M}$ to have negative real parts \cite{DeJesus}, which can be checked using the Routh-Hurwitz criterion. Once stability is established, the steady-state covariance matrix can be used to quantify quantum correlations such as entanglement and steering, as detailed in the following section.

Under the stability criterion, the Lyapunov equation in \eqref{eq:dlyap} is relaxed and the covariance matrix satisfies,
\begin{equation}\label{eq:lyap}
\mathbf{M}\mathbf{V} + \mathbf{V}\mathbf{M}^\top = -\mathbf{D}.
\end{equation}

\subsection{Quantification of quantum correlations}
\label{sec:quantification}

To quantify the entanglement between distinct modes $m$ and $n$ ($m, n \equiv a_1, a_2, b$) in the system, we employ the logarithmic negativity $E_{m|n}$, defined as~\cite{Horodecki2009},
\begin{equation}
E_{m|n} = \max \left[0, -\ln(2\nu_{m|n})\right],
\end{equation}
where
\begin{equation}
\nu_{m|n} = \frac{1}{\sqrt{2}} \sqrt{ \Sigma(V_{m|n}) - \sqrt{ \Sigma(V_{m|n})^2 - 4 \det V_{m|n} } },
\end{equation}
and $\Sigma(V_{m|n}) = \det V_m + \det V_n - 2 \det V_{m,n}$. Here, $V_{m|n}$ denotes the reduced covariance matrix for the modes of interest $m$ and $n$, given by,
\begin{equation}
V_{m|n} = \begin{pmatrix}
V_m & V_{m,n} \\
V_{m,n}^{\top} & V_n
\end{pmatrix},
\end{equation}
where $V_m$, $V_n$, and $V_{m,n}$ are $2\times2$ submatrices extracted from the full covariance matrix $V$. When the modes are entangled ($E_{m|n} > 0$), it is also possible to evaluate the directional influence of one mode on another, known as quantum steering.

Quantum steering from mode $m$ to mode $n$ and vice versa is quantified by~\cite{Uola2020,Xiang2022},
\begin{align}
    \mathcal{G}_{m \rightarrow n} &= \max \left[ 0, \frac{1}{2} \ln \frac{\det V_m}{4 \det V_{m|n}} \right], \\
    \mathcal{G}_{n \rightarrow m} &= \max \left[ 0, \frac{1}{2} \ln \frac{\det V_n}{4 \det V_{m|n}} \right].
\end{align}
If both $\mathcal{G}_{m \rightarrow n}$ and $\mathcal{G}_{n \rightarrow m}$ vanish, there is no steering between the modes. If only one is nonzero, the system exhibits one-way steering, and when both are nonzero, the system exhibits two-way steering. It should be highlighted that steering is generally asymmetric, and this asymmetry can be quantified by,
\begin{equation}
    \Delta \mathcal{G}_{m|n} = \left| \mathcal{G}_{m \rightarrow n} - \mathcal{G}_{n \rightarrow m} \right|.
\end{equation}
These metrics provide a comprehensive framework for quantifying quantum correlations generated in our system, as will be detailed later on.

\section{Enhancement of quantum correlations}\label{sec:entangl}

In this section, we investigate how saturable nonlinearities enable and control quantum correlations, i.e., entanglement and quantum steering, in our double-cavity optomechanical system. We focus on both the emergence and robustness of these correlations under experimentally realistic conditions. First, we analyze the generation and enhancement of bipartite entanglement; next, we investigate the  properties and asymmetry of quantum steering.

The numerical analysis is based on parameter regimes (see \Cref{tab:Param}) accessible in state-of-the-art experiments~\cite{Fang2025}, such as photonic crystal cavities~\cite{Monifi2016}, superconducting circuits~\cite{Brubaker2022}, and optomechanical systems \cite{Lake2020}. We operate in the red-sideband regime ($\Delta_j = \omega_m$), which is suitable for cooling and quantum state generation. We ensure that all parameters satisfy the Routh-Hurwitz criterion for linear stability~\cite{DeJesus}.

\begin{table}[htb]
\caption{Parameter used in our numerical simulations.}
\label{tab:Param}
    \begin{ruledtabular}
\begin{tabular}{lcl}
Parameter & Value & Description \\
\hline
$\omega_m$ & $1$--$10$~MHz & Mechanical frequency \\
$\kappa_j$ & $0.1$--$1$~MHz & Optical decay rate \\
$J$        & $0.1$--$0.5\,\omega_m$ & Photon hopping rate \\
$G_j$      & $0.2\,\omega_m$ & Optomechanical coupling \\
$n$        & $10$--$10^3$ & Thermal phonon number \\
\end{tabular}
    \end{ruledtabular}
\end{table}

The chosen parameters ensure both experimental feasibility and theoretical consistency, providing a solid foundation for the analysis of quantum correlations carried out in our proposal.

\subsection{Emergence of entanglement under saturable nonlinearity}

We begin by confirming that in the absence of saturable nonlinearity ($g_s = f_s = 0$), the system does not exhibit any steady-state quantum entanglement or steering. This establishes a clear baseline for assessing the impact of nonlinear effects.

\begin{figure*}[tbh]
  \centering
  \resizebox{\textwidth}{!}{
  \includegraphics{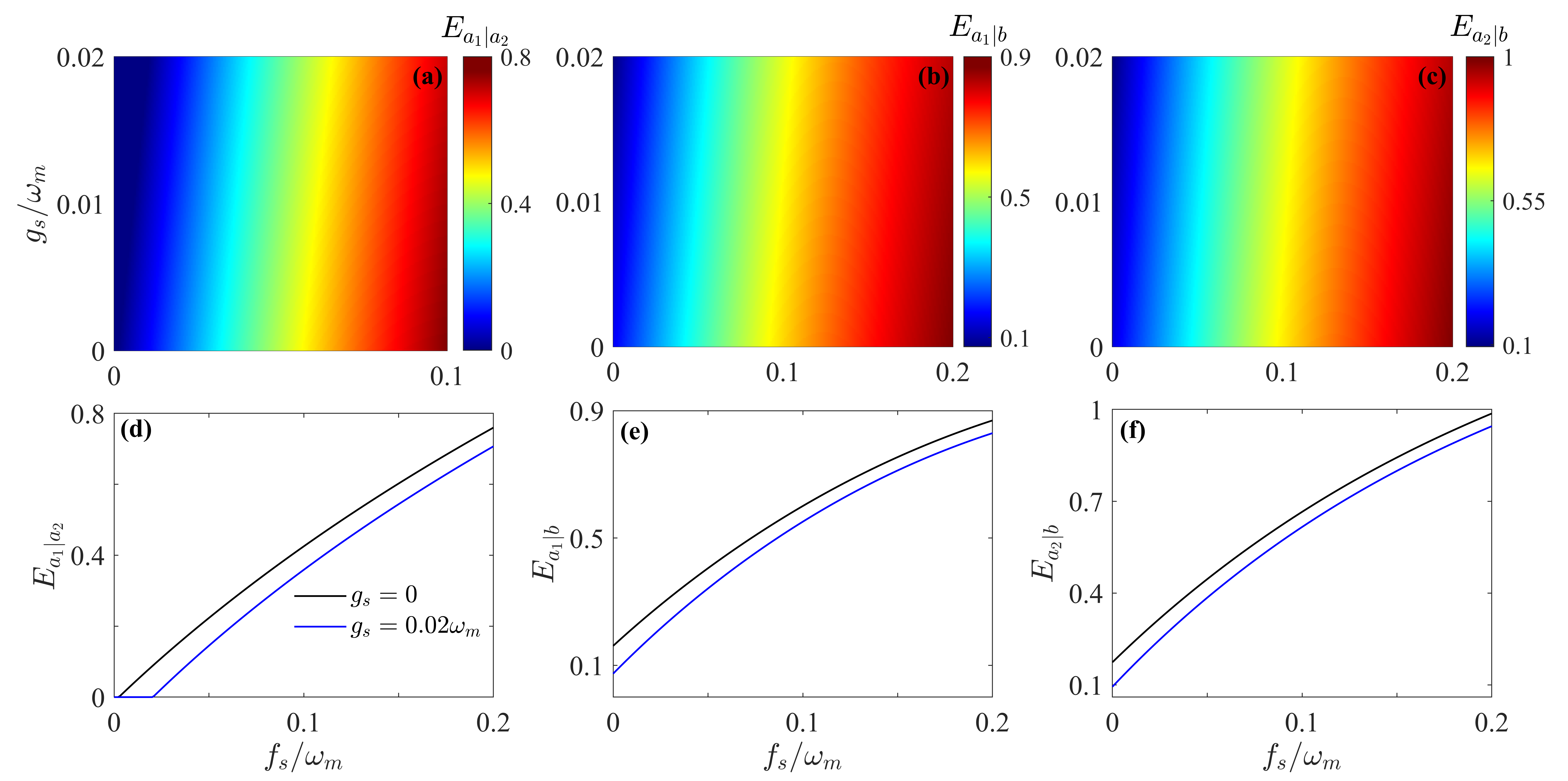}
  }
  \caption{Bipartite entanglement versus saturable gain $g_s / \omega_m$ and loss $f_s / \omega_m$. Panels (a--c) show 3D contour plots of (a) $E_{a_1|a_2}$, (b) $E_{a_1|b}$, and (c) $E_{a_2|b}$, with color scales indicating logarithmic negativity (0 to 0.5). Panels (d--f) show 2D slices at $g_s = 0$ for (d) $E_{a_1|a_2}$, (e) $E_{a_1|b}$, and (f) $E_{a_2|b}$. Parameters are $\kappa_j = 0.1 \omega_m$, $\gamma_m = 10^{-5} \omega_m$, $G_j = 0.2 \omega_m$, $J = 0.2 \omega_m$, $n = 100$, $\theta = \pi/2$, and $\Delta_j = \omega_m$.}
  \label{fig:Fig2}
\end{figure*}

\Cref{fig:Fig2} shows the logarithmic negativity for bipartite entanglement between optical modes ($E_{a_1|a_2}$) and optomechanical pairs ($E_{a_1|b}$, $E_{a_2|b}$), as defined in \Cref{sec:quantification}. The parameters used are $\kappa_j = 0.1 \omega_m$, $\gamma_m = 10^{-5} \omega_m$, $G_j = 0.2 \omega_m$, $J = 0.2 \omega_m$, $n = 100$, and $\theta = \pi/2$ (phase of optomechanical coupling).

It can be seen from \Cref{fig:Fig2}(a-c) that there is no entanglement without nonlinearities in the system ($g_s=0, f_s=0$). Activating saturable nonlinearity ($g_s, f_s \neq 0$) induces entanglement, with optomechanical pairs ($E_{a_1|b}$, $E_{a_2|b}$) exhibiting slightly stronger correlations than the purely optical pair ($E_{a_1|a_2}$). Notably, $E_{a_2|b}$ reaches values up to $0.5$ in logarithmic negativity, compared to $0.4$ for $E_{a_1|a_2}$ at $f_s = 0.1 \omega_m$ (see \Cref{fig:Fig2}). This enhancement arises from the nonlinear modulation of net decay rates ($g = g_s - \kappa_1$, $f = f_s + \kappa_2$), which alters the intracavity fields ($\alpha_j$) and effective optomechanical couplings ($G_j = g_j \alpha_j$). A key observation is the pronounced nonlinear asymmetry effect: saturable loss ($f_s$) is significantly more effective than gain ($g_s$) in generating and sustaining entanglement, as seen in \Cref{fig:Fig2}(d--f), where entanglement persists even when $g_s = 0$. This asymmetry originates from the ability of loss on stabilizing the system and enhancing the optomechanical interaction. In our model, saturable loss effectively increases the damping rate of the lossy cavity ($f = f_s + \kappa_2$), stabilizing the system by preventing noise amplification that typically occurs near parametric instability point.  This effect is particularly pronounced in optomechanical pairs involving the lossy cavity, where the loss-enhanced coupling $G_2$ triggers stronger correlations, as evidenced in \Cref{fig:Fig2}(c,f). These results demonstrate that saturable loss,  provides a robust and tunable mechanism for entanglement generation, even at elevated thermal occupations, paving the way for practical quantum information processing in optomechanical platforms~\cite{Fang2025}.

\subsection{Loss-induced enhancement and asymmetry in quantum steering}

We now turn to the analysis of quantum steering and its asymmetry under saturable nonlinearity. To elucidate the asymmetric nature of quantum correlations, we examine quantum steering ($\mathcal{G}_{m \to n}$) and its asymmetry ($\Delta \mathcal{G}_{m|n}$), with particular focus on regimes featuring dominant saturable loss ($f_s = 0.1\omega_m$) and weak gain ($g_s = 10^{-3}\omega_m$). This allows us to highlight the nonreciprocal effects induced by the nonlinear loss channel.

\begin{figure}[tbh]
  \centering
  \resizebox{0.5\textwidth}{!}{
  \includegraphics{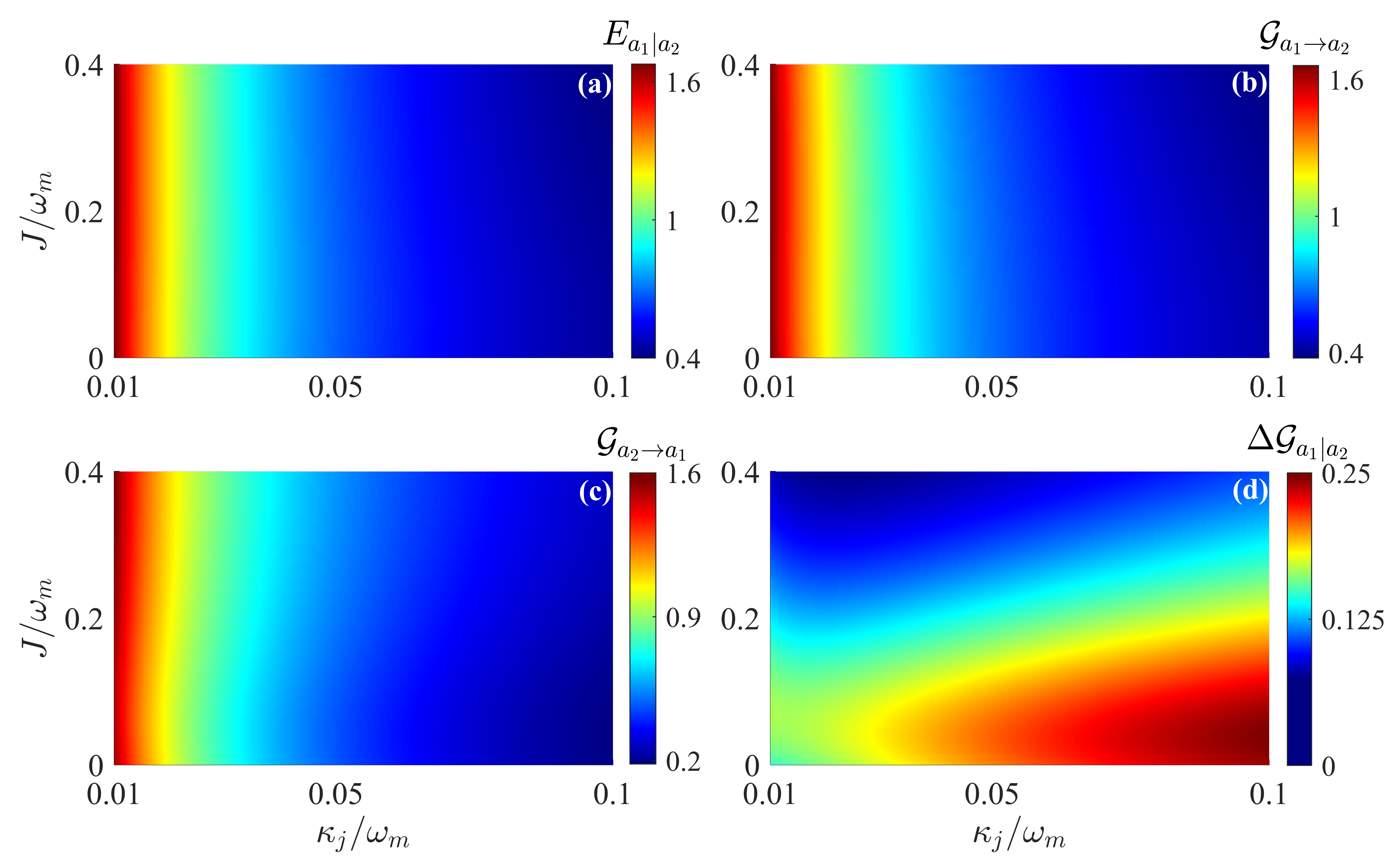}
  }
  \caption{Optical correlations versus optical decay rate $\kappa_j / \omega_m$ and photon hopping $J / \omega_m$. Panels show (a) entanglement $E_{a_1|a_2}$, (b) steering $\mathcal{G}_{a_1 \to a_2}$, (c) steering $\mathcal{G}_{a_2 \to a_1}$, and (d) asymmetry $\Delta \mathcal{G}_{a_1|a_2}$. Color scales indicate correlation strength (0 to 0.5 for $E$, 0 to 1 for $\mathcal{G}$, 0 to 0.5 for $\Delta \mathcal{G}$). Parameters are $g_s = 10^{-3} \omega_m$, $f_s = 0.1 \omega_m$, and as in \Cref{fig:Fig2}.}
  \label{fig:Fig3}
\end{figure}

\begin{figure}[tbh]
  \centering
  \resizebox{0.5\textwidth}{!}{
  \includegraphics{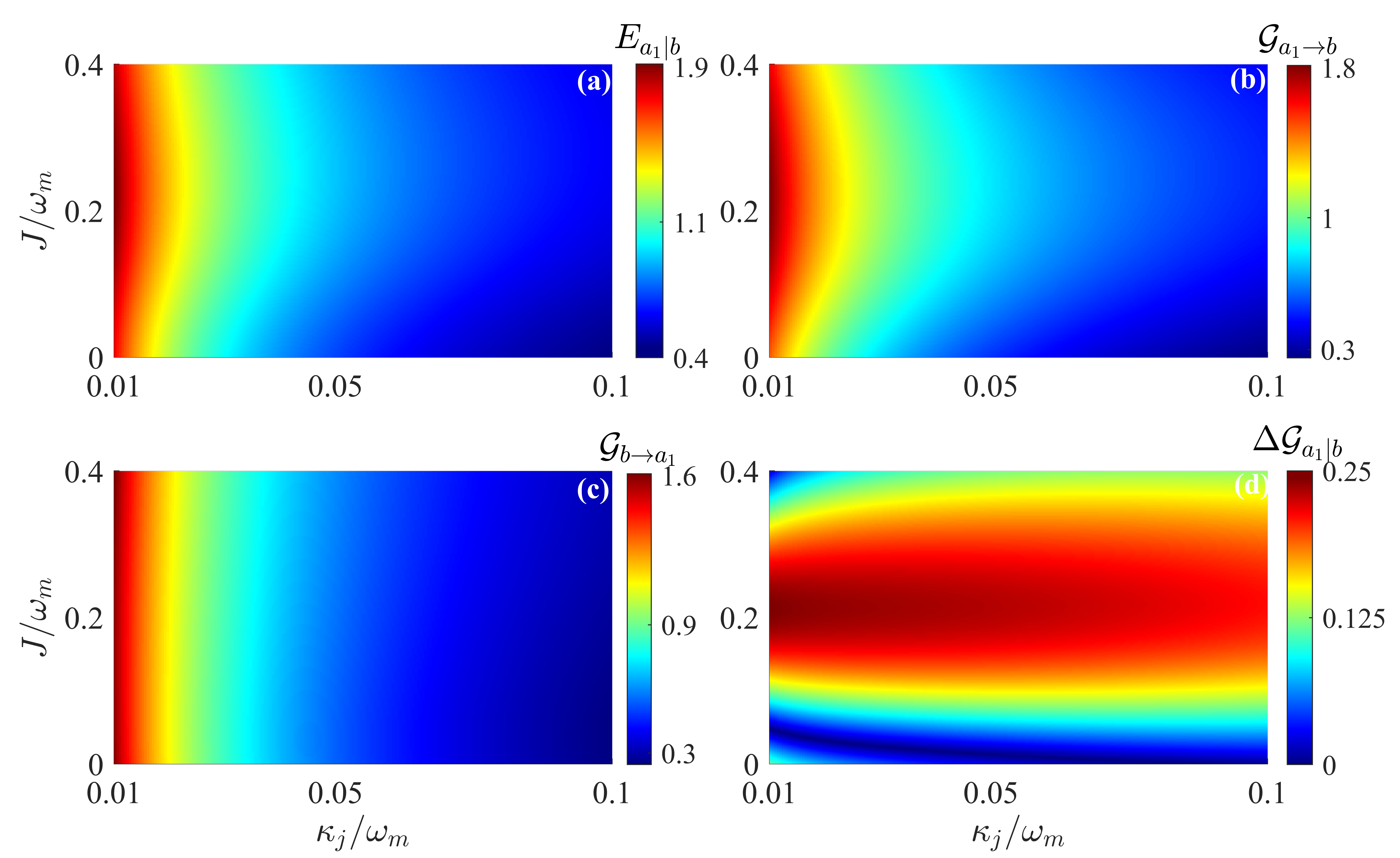}
  }
  \caption{Optomechanical correlations for $a_1|b$ versus $\kappa_j / \omega_m$ and $J / \omega_m$. Panels show (a) entanglement $E_{a_1|b}$, (b) steering $\mathcal{G}_{a_1 \to b}$, (c) steering $\mathcal{G}_{b \to a_1}$, and (d) asymmetry $\Delta \mathcal{G}_{a_1|b}$. Color scales are as in \Cref{fig:Fig3}. Parameters are $g_s = 10^{-3} \omega_m$, $f_s = 0.1 \omega_m$, and as in \Cref{fig:Fig2}.}
  \label{fig:Fig4}
\end{figure}

\begin{figure}[tbh]
  \centering
  \resizebox{0.5\textwidth}{!}{
  \includegraphics{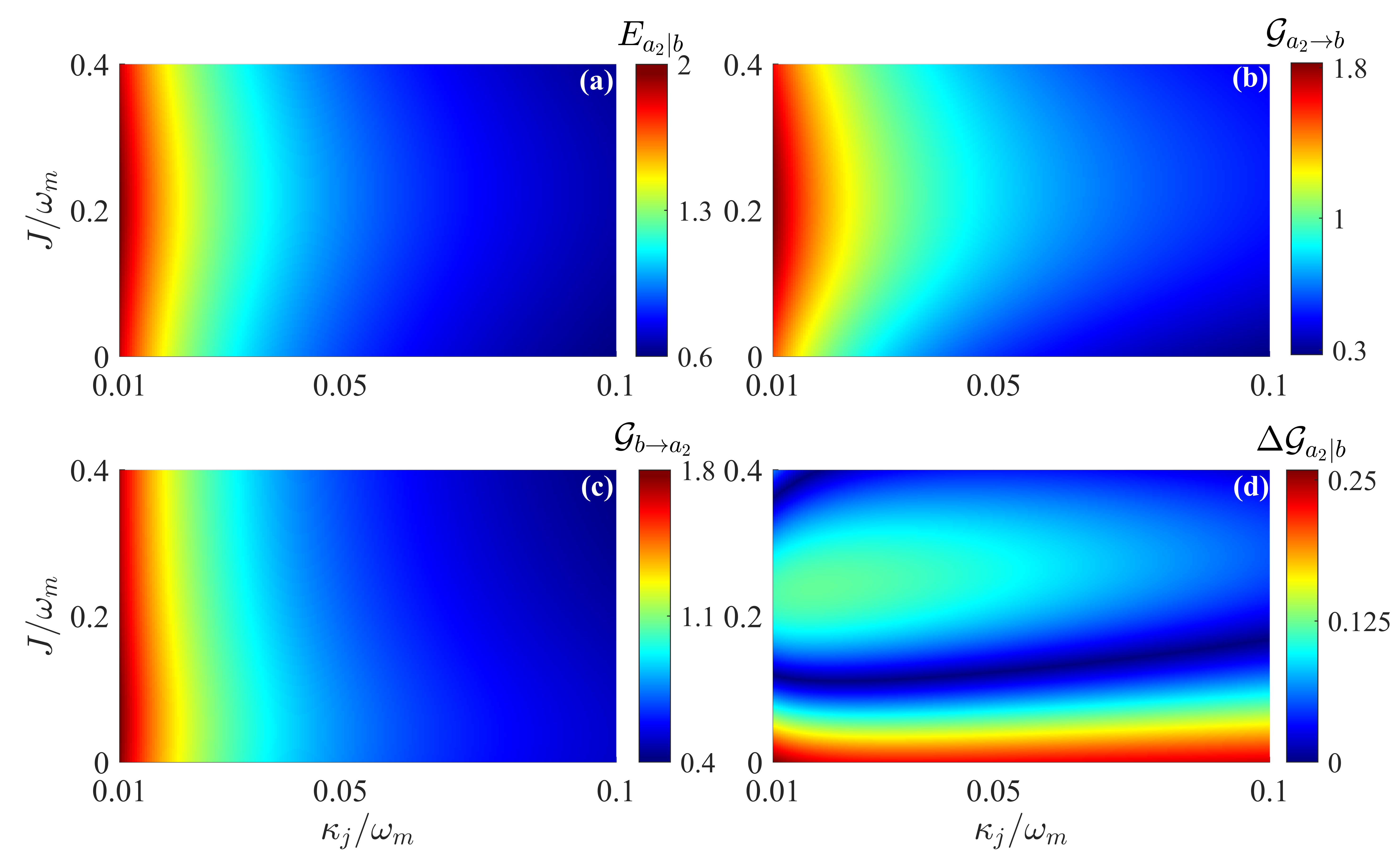}
  }
  \caption{Optomechanical correlations for $a_2|b$ versus $\kappa_j / \omega_m$ and $J / \omega_m$. Panels show (a) entanglement $E_{a_2|b}$, (b) steering $\mathcal{G}_{a_2 \to b}$, (c) steering $\mathcal{G}_{b \to a_2}$, and (d) asymmetry $\Delta \mathcal{G}_{a_2|b}$. Color scales are as in \Cref{fig:Fig3}. Parameters are $g_s = 10^{-3} \omega_m$, $f_s = 0.1 \omega_m$, and as in \Cref{fig:Fig2}.}
  \label{fig:Fig5}
\end{figure}

\Crefrange{fig:Fig3}{fig:Fig5} show that the strongest quantum correlations, including steering, emerge in the deep resolved-sideband regime ($\omega_m \gg \kappa_j$), where low optical decay rates suppress decoherence. The pronounced asymmetry in steering, especially for optomechanical pairs, reflects the nonreciprocal influence of saturable loss on the system dynamics and reveals the potential for quantum information transfer.

\Cref{fig:Fig3} presents the optical correlations ($a_1|a_2$), where two-way steering ($\mathcal{G}_{a_1 \to a_2}$ and $\mathcal{G}_{a_2 \to a_1} > 0$) is observed, with moderate asymmetry ($\Delta \mathcal{G}_{a_1|a_2}$) peaking at $J \approx 0.2 \omega_m$. In \Cref{fig:Fig4}, the optomechanical correlations for $a_1|b$ exhibit slightly higher entanglement and steering than in the purely optical case, a consequence of the non-linear modulation of the effective gain ($g = g_s - \kappa_1$). \Cref{fig:Fig5} displays the highest correlations for $a_2|b$, attributed to the saturable loss ($f_s$) that increases the effective damping ($f = f_s + \kappa_2$), which stabilizes $\alpha_2$ and improves the optomechanical coupling $G_2$.
 
\begin{figure*}[tbh]
  \centering
  \resizebox{\textwidth}{!}{
  \includegraphics{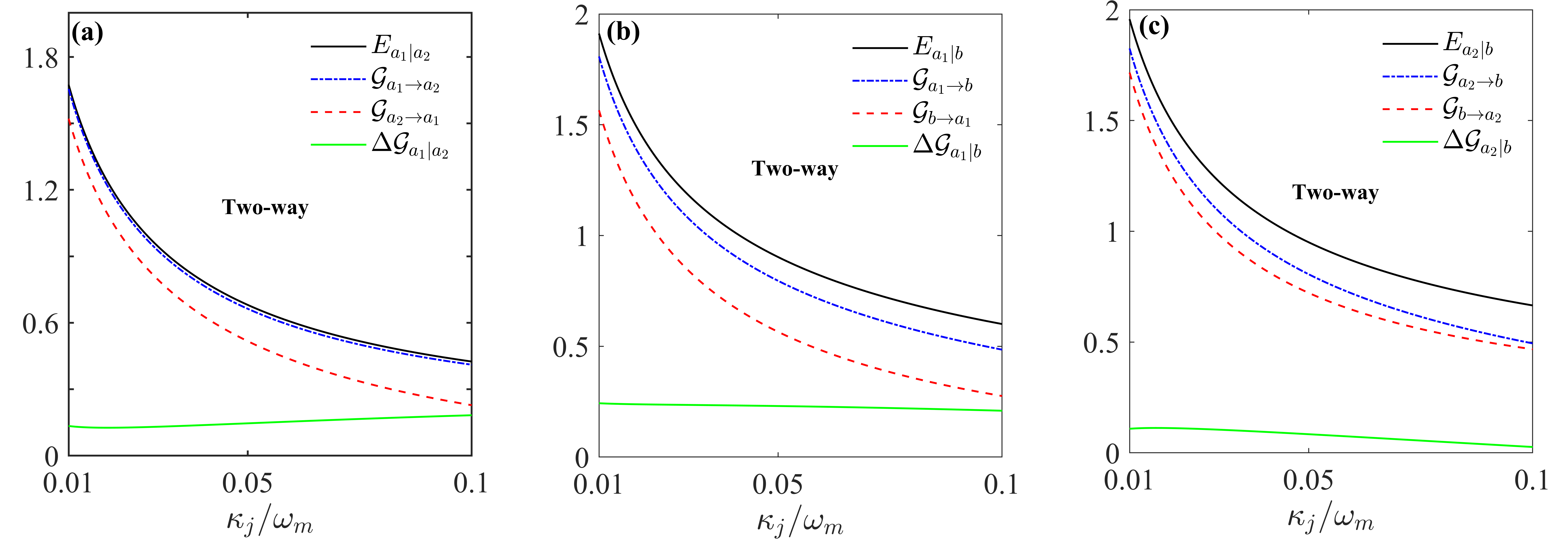}
  }
  \caption{Quantum correlations versus optical decay rate $\kappa_j / \omega_m$ at $J = 0.2 \omega_m$. Panels show (a) optical correlations ($E_{a_1|a_2}$, $\mathcal{G}_{a_1 \to a_2}$, $\mathcal{G}_{a_2 \to a_1}$), (b) optomechanical correlations for $a_1|b$ ($E_{a_1|b}$, $\mathcal{G}_{a_1 \to b}$, $\mathcal{G}_{b \to a_1}$), and (c) optomechanical correlations for $a_2|b$ ($E_{a_2|b}$, $\mathcal{G}_{a_2 \to b}$, $\mathcal{G}_{b \to a_2}$). Solid lines denote entanglement, dashed lines denote optical-to-optical/mechanical steering, and dotted lines denote mechanical-to-optical steering. Parameters are $g_s = 10^{-3} \omega_m$, $f_s = 0.1 \omega_m$, and as in \Cref{fig:Fig2}.}
  \label{fig:Fig6}
\end{figure*}

A key feature, shown in \Cref{fig:Fig6}, is that steering from optical to mechanical modes ($\mathcal{G}_{a_j \to b}$) is consistently stronger than the reverse ($\mathcal{G}_{b \to a_j}$). The 2D slices at $J = 0.2 \omega_m$ confirm that $\mathcal{G}_{a_1 \to b}$ and $\mathcal{G}_{a_2 \to b}$ dominate over their mechanical-to-optical counterparts. This asymmetry, rooted in saturable loss, enables robust quantum control and is highly advantageous for quantum tasks.

The pronounced asymmetry in quantum steering suggests promising applications in nonreciprocal quantum information processing. For example, strong optical-to-mechanical steering ($\mathcal{G}_{a_j \to b} > \mathcal{G}_{b \to a_j}$) could be exploited for one-way quantum state transfer or for the implementation of directional quantum gates, where information flows predominantly from the optical to the mechanical degree of freedom. Such nonreciprocal behavior is particularly valuable in quantum networks, where controlling the direction of information flow is essential for quantum communication and distributed quantum computing~\cite{Slussarenko2019}. These capabilities align with the robust correlation generation demonstrated in our analysis, further enhancing the system’s utility for advanced quantum information protocols. Overall, these findings reveal the central role of saturable nonlinearity in engineering robust and directional quantum correlations, with significant implications for the development of practical quantum information technologies.

\subsection{Stability and covariance analysis}

The stability of our nonlinear optomechanical system is fundamental for the reliable generation of quantum correlations. As detailed in \Cref{sec:model}, the quadrature dynamics are governed by the drift matrix $\rm{M}$ (Eq.~\eqref{eq:matrix}), whose eigenvalues must possess negative real parts to ensure dynamical stability. We rigorously enforce this requirement by combining the Routh-Hurwitz criterion with the Lyapunov covariance analysis~\cite{DeJesus}. For the experimentally relevant parameter regime ($\kappa_j = 0.1\omega_m$, $G_j = 0.2\omega_m$, $J = 0.2\omega_m$), the Routh-Hurwitz conditions confirm the stability in all the saturable gain/loss combinations explored in \Crefrange{fig:Fig2}{fig:Fig5}. This ensures that the subsequent analysis of quantum correlations is performed exclusively in dynamically stable regimes.

The steady-state covariance matrix $\rm{V}$ is then determined by solving the Lyapunov equation, $\mathbf{M}\mathbf{V} + \mathbf{V}\mathbf{M}^\top = -\mathbf{D}$ (Eq.~\eqref{eq:lyap}), where $\rm{D}$ encodes both thermal and optical noise. This approach extends conventional optomechanical stability analyses~\cite{Bemani2019} by explicitly incorporating the nonlinear gain/loss modulation described in Eqs.~\eqref{eq:sat}. Our stability framework reveals two key advantages: (i) saturable loss increases the effective decay rate $f = f_s + \kappa_2$, thereby delaying parametric instabilities in our system; and (ii) the nonlinear gain/loss balance maintains $\Re[\lambda(\mathbf{M})] < 0$ even at elevated thermal occupations ($n \leq 10^3$). These features underpin the robust correlation generation demonstrated in subsequent sections and highlight the suitability of our scheme for practical quantum technologies.

To further illustrate the dependence of quantum correlations on key system parameters, \Cref{tab:parameter_dependence} summarizes the observed trends and associated stability considerations.

\begin{table}[htb]
\caption{Parameter dependence of quantum correlations.}
\label{tab:parameter_dependence}
\centering
    \begin{ruledtabular}
\begin{tabular}{l|c|p{2cm}|p{2cm}}
Parameter & Range & Impact on correlations & Stability notes \\
\hline
$J$ (hopping) & $0.1$--$0.5\,\omega_m$ & Increases up to $0.3\,\omega_m$, then saturates & Stable for $J<0.4\,\omega_m$ \\\hline
$f_s$ (loss) & $0$--$0.1\,\omega_m$ & Linear increase in entanglement & Enhances stability near EPs \\\hline
$n$ (phonon) & $10$--$10^3$ & Decreases slowly, robust up to $n=10^3$ & Key for room-temperature operation \\
\end{tabular}
    \end{ruledtabular}
\end{table}

\subsection{Thermal noise resilience}

We now examine the robustness of the generated quantum correlations against thermal noise, a key requirement for practical quantum technologies operating at room temperatures.

\begin{figure*}[tbh]
  \centering
  \resizebox{\textwidth}{!}{
  \includegraphics{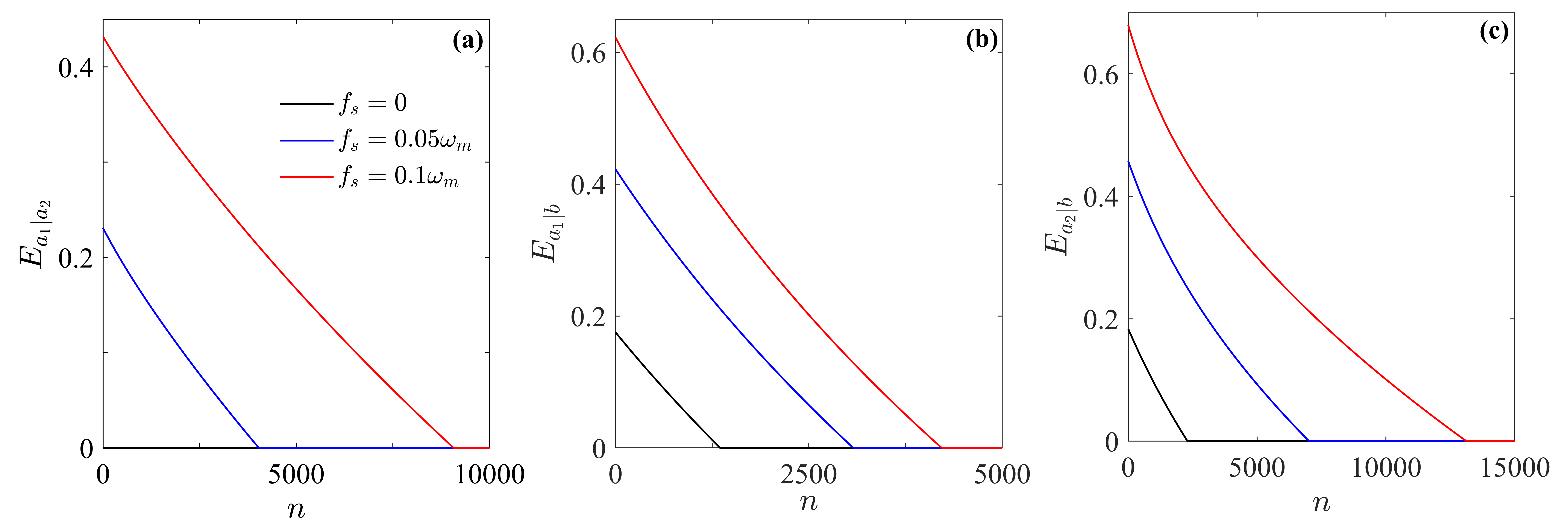}
  }
  \caption{Thermal robustness of bipartite entanglement. Logarithmic negativity versus thermal phonon number $n$ for fixed $g_s = 10^{-3}\omega_m$ and varying saturable loss: $f_s = 0$ (solid), $0.05\omega_m$ (dashed), $0.1\omega_m$ (dotted). (a) $E_{a_1|a_2}$, (b) $E_{a_1|b}$, (c) $E_{a_2|b}$. Other parameters: $\kappa_j = 0.1\omega_m$, $\gamma_m = 10^{-5}\omega_m$, $G_j = 0.2\omega_m$, $J = 0.2\omega_m$, $\theta = \pi/2$, $\Delta_j = \omega_m$.}
  \label{fig:Fig7}
\end{figure*}

\Cref{fig:Fig7} shows the logarithmic negativity as a function of the thermal phonon number $n$, using the same parameters as in \Cref{fig:Fig2} with $g_s = 10^{-3}\omega_m$, $f_s = 0,\, 0.05,\, 0.1\,\omega_m$, and $J = 0.2\,\omega_m$. The results in \Cref{fig:Fig7} display a remarkable resilience of the entanglement as thermal noise increases, with quantum correlations persisting up to $n \approx 10^3$ (corresponding to $T \approx 300\,\text{K}$ for $\omega_m \approx 1\,\text{MHz}$). Two key features are observed: (i) saturable loss enhances thermal tolerance by 2--3 orders of magnitude compared to conventional optomechanical systems (compare curves for $f_s=0$ and $f_s=0.1\omega_m$); (ii) the $a_2|b$ pair exhibits the strongest robustness, attributable to loss-stabilized coupling $G_2$. This enhanced thermal resilience arises from two mechanisms: (i) an increase of the effective loss rate $f$ that suppresses the injection of mechanical noise by modifying the diffusion matrix $\mathbf{D}$, and (ii) the non-linear saturation that prevents gain-induced parametric amplification. As a result, the system maintains large amount of quantum correlations at high thermal phonon number, meeting threshold requirements for room temperature quantum information processing and enabling robust operation in quantum networks and sensing applications~\cite{Slussarenko2019, Xia2023}.

\subsection{Experimental realization}

To validate our theoretical predictions, we propose an experimental implementation using photonic crystal cavities, as demonstrated in~\cite{Monifi2016}. Similarly, our proposal can be also implemented in optomechanical systems as implemented in \cite{Lake2020}.  These platforms are ideally suited for engineering the required saturable nonlinearities and for achieving the parameter regimes explored in this work. In such systems, saturable gain can be realized via embedded quantum dots or other nonlinear media, while saturable loss can be introduced through controlled absorption mechanisms. The two optical cavities are coupled via evanescent fields, and the mechanical resonator can be implemented as a suspended membrane or a nanobeam integrated within the photonic crystal structure.

Quantum correlations can be measured using homodyne detection for the optical fields and displacement (or heterodyne) measurements for the mechanical mode. These measurements enable reconstruction of the entire covariance matrix, allowing direct verification of entanglement and steering as discussed in Section~\ref{sec:quantification}. Potential challenges, such as thermal noise and detection inefficiencies, can be mitigated by advanced calibration and noise reduction techniques. The parameter regime considered in our work ($\kappa_j = 0.1 \omega_m$, $G_j = 0.2 \omega_m$, $J = 0.2 \omega_m$) is compatible with current state-of-the-art experimental capabilities.

\section{Conclusion} \label{sec:concl}

In this work, we have theoretically investigated the generation and control of quantum correlations in a double-cavity optomechanical system, where two optical modes, coupled via photon hopping ($J$), interact with a common mechanical resonator. By introducing saturable nonlinearities that induce controllable gain ($g_s$) and loss ($f_s$) in the optical cavities, we generated robust bipartite entanglement and two-way quantum steering between optical and optomechanical modes. Our key findings reveal that: (i) quantum correlations, i.e., bipartite entanglement and two-way steering are induced in the system via saturable nonlinearities, and (ii) the saturated loss is mainly responsible for the enhancement of the quantum correlations. Moreover, the generated quantum correlations exhibit strong resilience to thermal noise, with entanglement persisting up to $n \approx 10^3$ (corresponding to room temperature, $T \approx 300\,\text{K}$ for $\omega_m \approx 1\,\text{MHz}$), surpassing conventional optomechanical systems by two to three orders of magnitude. Our results suggest saturable nonlinearity as a versatile tool to engineer robust quantum correlations in optomechanical systems, with promising applications in quantum information processing, communication, and sensing.

\section*{Acknowledgments}

P.D. acknowledges the Iso-Lomso Fellowship at Stellenbosch Institute for Advanced Study (STIAS), Wallenberg Research Centre at Stellenbosch University, Stellenbosch 7600, South Africa, and The Institute for Advanced Study, Wissenschaftskolleg zu Berlin, Wallotstrasse 19, 14193 Berlin, Germany. The authors extend their appreciation to Northern Border University, Saudi Arabia, for supporting this work through project number (NBU-CRP-2025-3021). This work was supported by Princess Nourah bint Abdulrahman University Researchers Supporting Project number (PNURSP2025R399), Princess Nourah bint Abdulrahman University, Riyadh, Saudi Arabia. The authors are thankful to the Deanship of Graduate Studies and Scientific Research at University of Bisha for supporting this work through the Fast-Track Research Support Program. 

\textbf{Author Contributions:} D.R.K.M. and P.D. conceptualized the work and carried out the simulations and analysis. E.K.B., A.-H. A.-A. and M.R.E. participated in all the discussions and provided useful methodology and suggestions for the final version of the manuscript. R.A. and S.G.N.E. participated in the discussions and supervised the work. All authors participated equally in the writing, discussions, and the preparation of the final version of the manuscript.

\textbf{Competing Interests:} All authors declare no competing interests.

\textbf{Data Availability:}
Relevant data are included in the manuscript and supporting information. Supplement data are available upon reasonable request.

\newpage

\bibliography{Gain_saturation} 

\begin{thebibliography}{37}%
\makeatletter
\providecommand \@ifxundefined [1]{%
 \@ifx{#1\undefined}
}%
\providecommand \@ifnum [1]{%
 \ifnum #1\expandafter \@firstoftwo
 \else \expandafter \@secondoftwo
 \fi
}%
\providecommand \@ifx [1]{%
 \ifx #1\expandafter \@firstoftwo
 \else \expandafter \@secondoftwo
 \fi
}%
\providecommand \natexlab [1]{#1}%
\providecommand \enquote  [1]{``#1''}%
\providecommand \bibnamefont  [1]{#1}%
\providecommand \bibfnamefont [1]{#1}%
\providecommand \citenamefont [1]{#1}%
\providecommand \href@noop [0]{\@secondoftwo}%
\providecommand \href [0]{\begingroup \@sanitize@url \@href}%
\providecommand \@href[1]{\@@startlink{#1}\@@href}%
\providecommand \@@href[1]{\endgroup#1\@@endlink}%
\providecommand \@sanitize@url [0]{\catcode `\\12\catcode `\$12\catcode
  `\&12\catcode `\#12\catcode `\^12\catcode `\_12\catcode `\%12\relax}%
\providecommand \@@startlink[1]{}%
\providecommand \@@endlink[0]{}%
\providecommand \url  [0]{\begingroup\@sanitize@url \@url }%
\providecommand \@url [1]{\endgroup\@href {#1}{\urlprefix }}%
\providecommand \urlprefix  [0]{URL }%
\providecommand \Eprint [0]{\href }%
\providecommand \doibase [0]{https://doi.org/}%
\providecommand \selectlanguage [0]{\@gobble}%
\providecommand \bibinfo  [0]{\@secondoftwo}%
\providecommand \bibfield  [0]{\@secondoftwo}%
\providecommand \translation [1]{[#1]}%
\providecommand \BibitemOpen [0]{}%
\providecommand \bibitemStop [0]{}%
\providecommand \bibitemNoStop [0]{.\EOS\space}%
\providecommand \EOS [0]{\spacefactor3000\relax}%
\providecommand \BibitemShut  [1]{\csname bibitem#1\endcsname}%
\let\auto@bib@innerbib\@empty
\bibitem [{\citenamefont {Monifi}\ \emph {et~al.}(2016)\citenamefont {Monifi},
  \citenamefont {Zhang}, \citenamefont {Ozdemir}, \citenamefont {Peng},
  \citenamefont {Liu}, \citenamefont {Bo}, \citenamefont {Nori},\ and\
  \citenamefont {Yang}}]{Monifi2016}%
  \BibitemOpen
  \bibfield  {author} {\bibinfo {author} {\bibfnamefont {F.}~\bibnamefont
  {Monifi}}, \bibinfo {author} {\bibfnamefont {J.}~\bibnamefont {Zhang}},
  \bibinfo {author} {\bibfnamefont {S.~K.}\ \bibnamefont {Ozdemir}}, \bibinfo
  {author} {\bibfnamefont {B.}~\bibnamefont {Peng}}, \bibinfo {author}
  {\bibfnamefont {Y.-x.}\ \bibnamefont {Liu}}, \bibinfo {author} {\bibfnamefont
  {F.}~\bibnamefont {Bo}}, \bibinfo {author} {\bibfnamefont {F.}~\bibnamefont
  {Nori}},\ and\ \bibinfo {author} {\bibfnamefont {L.}~\bibnamefont {Yang}},\
  }\bibfield  {title} {\bibinfo {title} {Optomechanically induced stochastic
  resonance and chaos transfer between optical fields},\ }\href
  {https://doi.org/10.1038/nphoton.2016.73} {\bibfield  {journal} {\bibinfo
  {journal} {Nature Photonics}\ }\textbf {\bibinfo {volume} {10}},\ \bibinfo
  {pages} {399} (\bibinfo {year} {2016})}\BibitemShut {NoStop}%
\bibitem [{\citenamefont {Navarro-Urrios}\ \emph {et~al.}(2017)\citenamefont
  {Navarro-Urrios}, \citenamefont {Capuj}, \citenamefont {Colombano},
  \citenamefont {García}, \citenamefont {Sledzinska}, \citenamefont {Alzina},
  \citenamefont {Griol}, \citenamefont {Martínez},\ and\ \citenamefont
  {Sotomayor-Torres}}]{Navarro2017}%
  \BibitemOpen
  \bibfield  {author} {\bibinfo {author} {\bibfnamefont {D.}~\bibnamefont
  {Navarro-Urrios}}, \bibinfo {author} {\bibfnamefont {N.~E.}\ \bibnamefont
  {Capuj}}, \bibinfo {author} {\bibfnamefont {M.~F.}\ \bibnamefont
  {Colombano}}, \bibinfo {author} {\bibfnamefont {P.~D.}\ \bibnamefont
  {García}}, \bibinfo {author} {\bibfnamefont {M.}~\bibnamefont {Sledzinska}},
  \bibinfo {author} {\bibfnamefont {F.}~\bibnamefont {Alzina}}, \bibinfo
  {author} {\bibfnamefont {A.}~\bibnamefont {Griol}}, \bibinfo {author}
  {\bibfnamefont {A.}~\bibnamefont {Martínez}},\ and\ \bibinfo {author}
  {\bibfnamefont {C.~M.}\ \bibnamefont {Sotomayor-Torres}},\ }\bibfield
  {title} {\bibinfo {title} {Nonlinear dynamics and chaos in an optomechanical
  beam},\ }\href {https://doi.org/10.1038/ncomms14965} {\bibfield  {journal}
  {\bibinfo  {journal} {Nature Communications}\ }\textbf {\bibinfo {volume}
  {8}},\ \bibinfo {pages} {14965} (\bibinfo {year} {2017})}\BibitemShut
  {NoStop}%
\bibitem [{\citenamefont {Mbokop~Tchounda}\ \emph {et~al.}(2024)\citenamefont
  {Mbokop~Tchounda}, \citenamefont {Djorwé}, \citenamefont {Tchakui},\ and\
  \citenamefont {Nana~Engo}}]{Mbok2024}%
  \BibitemOpen
  \bibfield  {author} {\bibinfo {author} {\bibfnamefont {S.~R.}\ \bibnamefont
  {Mbokop~Tchounda}}, \bibinfo {author} {\bibfnamefont {P.}~\bibnamefont
  {Djorwé}}, \bibinfo {author} {\bibfnamefont {M.~V.}\ \bibnamefont
  {Tchakui}},\ and\ \bibinfo {author} {\bibfnamefont {S.~G.}\ \bibnamefont
  {Nana~Engo}},\ }\bibfield  {title} {\bibinfo {title} {Chaos control and
  exceptional point engineering via dissipative optomechanical coupling},\
  }\href {https://doi.org/10.1088/1402-4896/ad195c} {\bibfield  {journal}
  {\bibinfo  {journal} {Physica Scripta}\ }\textbf {\bibinfo {volume} {99}},\
  \bibinfo {pages} {025215} (\bibinfo {year} {2024})}\BibitemShut {NoStop}%
\bibitem [{\citenamefont {Sheng}\ \emph {et~al.}(2020)\citenamefont {Sheng},
  \citenamefont {Wei}, \citenamefont {Yang},\ and\ \citenamefont
  {Wu}}]{Sheng2020}%
  \BibitemOpen
  \bibfield  {author} {\bibinfo {author} {\bibfnamefont {J.}~\bibnamefont
  {Sheng}}, \bibinfo {author} {\bibfnamefont {X.}~\bibnamefont {Wei}}, \bibinfo
  {author} {\bibfnamefont {C.}~\bibnamefont {Yang}},\ and\ \bibinfo {author}
  {\bibfnamefont {H.}~\bibnamefont {Wu}},\ }\bibfield  {title} {\bibinfo
  {title} {Self-organized synchronization of phonon lasers},\ }\href
  {https://doi.org/10.1103/PhysRevLett.124.053604} {\bibfield  {journal}
  {\bibinfo  {journal} {Physical Review Letters}\ }\textbf {\bibinfo {volume}
  {124}},\ \bibinfo {pages} {053604} (\bibinfo {year} {2020})}\BibitemShut
  {NoStop}%
\bibitem [{\citenamefont {Djorwé}\ \emph {et~al.}(2020)\citenamefont
  {Djorwé}, \citenamefont {Pennec},\ and\ \citenamefont
  {Djafari-Rouhani}}]{Djor2020}%
  \BibitemOpen
  \bibfield  {author} {\bibinfo {author} {\bibfnamefont {P.}~\bibnamefont
  {Djorwé}}, \bibinfo {author} {\bibfnamefont {Y.}~\bibnamefont {Pennec}},\
  and\ \bibinfo {author} {\bibfnamefont {B.}~\bibnamefont {Djafari-Rouhani}},\
  }\bibfield  {title} {\bibinfo {title} {Self-organized synchronization of
  mechanically coupled resonators based on optomechanics gain-loss balance},\
  }\href {https://doi.org/10.1103/PhysRevB.102.155410} {\bibfield  {journal}
  {\bibinfo  {journal} {Physical Review B}\ }\textbf {\bibinfo {volume}
  {102}},\ \bibinfo {pages} {155410} (\bibinfo {year} {2020})}\BibitemShut
  {NoStop}%
\bibitem [{\citenamefont {Rodrigues}\ \emph {et~al.}(2021)\citenamefont
  {Rodrigues}, \citenamefont {Kersul}, \citenamefont {Primo}, \citenamefont
  {Lipson}, \citenamefont {Alegre},\ and\ \citenamefont
  {Wiederhecker}}]{Rodrigues2021}%
  \BibitemOpen
  \bibfield  {author} {\bibinfo {author} {\bibfnamefont {C.~C.}\ \bibnamefont
  {Rodrigues}}, \bibinfo {author} {\bibfnamefont {C.~M.}\ \bibnamefont
  {Kersul}}, \bibinfo {author} {\bibfnamefont {A.~G.}\ \bibnamefont {Primo}},
  \bibinfo {author} {\bibfnamefont {M.}~\bibnamefont {Lipson}}, \bibinfo
  {author} {\bibfnamefont {T.~P.~M.}\ \bibnamefont {Alegre}},\ and\ \bibinfo
  {author} {\bibfnamefont {G.~S.}\ \bibnamefont {Wiederhecker}},\ }\bibfield
  {title} {\bibinfo {title} {Optomechanical synchronization across multi-octave
  frequency spans},\ }\href {https://doi.org/10.1038/s41467-021-25884-x}
  {\bibfield  {journal} {\bibinfo  {journal} {Nature Communications}\ }\textbf
  {\bibinfo {volume} {12}},\ \bibinfo {pages} {5625} (\bibinfo {year}
  {2021})}\BibitemShut {NoStop}%
\bibitem [{\citenamefont {Djorwe}\ \emph {et~al.}(2019)\citenamefont {Djorwe},
  \citenamefont {Pennec},\ and\ \citenamefont {Djafari-Rouhani}}]{Djorwe2019}%
  \BibitemOpen
  \bibfield  {author} {\bibinfo {author} {\bibfnamefont {P.}~\bibnamefont
  {Djorwe}}, \bibinfo {author} {\bibfnamefont {Y.}~\bibnamefont {Pennec}},\
  and\ \bibinfo {author} {\bibfnamefont {B.}~\bibnamefont {Djafari-Rouhani}},\
  }\bibfield  {title} {\bibinfo {title} {Exceptional point enhances sensitivity
  of optomechanical mass sensors},\ }\href
  {https://doi.org/10.1103/physrevapplied.12.024002} {\bibfield  {journal}
  {\bibinfo  {journal} {Physical Review Applied}\ }\textbf {\bibinfo {volume}
  {12}},\ \bibinfo {pages} {024002} (\bibinfo {year} {2019})}\BibitemShut
  {NoStop}%
\bibitem [{\citenamefont {Li}\ \emph {et~al.}(2021)\citenamefont {Li},
  \citenamefont {Ou}, \citenamefont {Lei},\ and\ \citenamefont {Liu}}]{Li2021}%
  \BibitemOpen
  \bibfield  {author} {\bibinfo {author} {\bibfnamefont {B.-B.}\ \bibnamefont
  {Li}}, \bibinfo {author} {\bibfnamefont {L.}~\bibnamefont {Ou}}, \bibinfo
  {author} {\bibfnamefont {Y.}~\bibnamefont {Lei}},\ and\ \bibinfo {author}
  {\bibfnamefont {Y.-C.}\ \bibnamefont {Liu}},\ }\bibfield  {title} {\bibinfo
  {title} {Cavity optomechanical sensing},\ }\href
  {https://doi.org/10.1515/nanoph-2021-0256} {\bibfield  {journal} {\bibinfo
  {journal} {Nanophotonics}\ }\textbf {\bibinfo {volume} {10}},\ \bibinfo
  {pages} {2799} (\bibinfo {year} {2021})}\BibitemShut {NoStop}%
\bibitem [{\citenamefont {Yan}\ \emph {et~al.}(2023)\citenamefont {Yan},
  \citenamefont {He},\ and\ \citenamefont {Lin}}]{Yan2023}%
  \BibitemOpen
  \bibfield  {author} {\bibinfo {author} {\bibfnamefont {Z.~F.}\ \bibnamefont
  {Yan}}, \bibinfo {author} {\bibfnamefont {B.}~\bibnamefont {He}},\ and\
  \bibinfo {author} {\bibfnamefont {Q.}~\bibnamefont {Lin}},\ }\bibfield
  {title} {\bibinfo {title} {Force sensing with an optomechanical system at
  room temperature},\ }\href {https://doi.org/10.1103/PhysRevA.107.013529}
  {\bibfield  {journal} {\bibinfo  {journal} {Physical Review A}\ }\textbf
  {\bibinfo {volume} {107}},\ \bibinfo {pages} {013529} (\bibinfo {year}
  {2023})}\BibitemShut {NoStop}%
\bibitem [{\citenamefont {Djorwé}\ \emph {et~al.}(2024)\citenamefont
  {Djorwé}, \citenamefont {Asjad}, \citenamefont {Pennec}, \citenamefont
  {Dutykh},\ and\ \citenamefont {Djafari-Rouhani}}]{Dj2024}%
  \BibitemOpen
  \bibfield  {author} {\bibinfo {author} {\bibfnamefont {P.}~\bibnamefont
  {Djorwé}}, \bibinfo {author} {\bibfnamefont {M.}~\bibnamefont {Asjad}},
  \bibinfo {author} {\bibfnamefont {Y.}~\bibnamefont {Pennec}}, \bibinfo
  {author} {\bibfnamefont {D.}~\bibnamefont {Dutykh}},\ and\ \bibinfo {author}
  {\bibfnamefont {B.}~\bibnamefont {Djafari-Rouhani}},\ }\bibfield  {title}
  {\bibinfo {title} {Parametrically enhancing sensor sensitivity at an
  exceptional point},\ }\href
  {https://doi.org/10.1103/PhysRevResearch.6.033284} {\bibfield  {journal}
  {\bibinfo  {journal} {Physical Review Research}\ }\textbf {\bibinfo {volume}
  {6}},\ \bibinfo {pages} {033284} (\bibinfo {year} {2024})}\BibitemShut
  {NoStop}%
\bibitem [{\citenamefont {Brubaker}\ \emph {et~al.}(2022)\citenamefont
  {Brubaker}, \citenamefont {Kindem}, \citenamefont {Urmey}, \citenamefont
  {Mittal}, \citenamefont {Delaney}, \citenamefont {Burns}, \citenamefont
  {Vissers}, \citenamefont {Lehnert},\ and\ \citenamefont
  {Regal}}]{Brubaker2022}%
  \BibitemOpen
  \bibfield  {author} {\bibinfo {author} {\bibfnamefont {B.}~\bibnamefont
  {Brubaker}}, \bibinfo {author} {\bibfnamefont {J.}~\bibnamefont {Kindem}},
  \bibinfo {author} {\bibfnamefont {M.}~\bibnamefont {Urmey}}, \bibinfo
  {author} {\bibfnamefont {S.}~\bibnamefont {Mittal}}, \bibinfo {author}
  {\bibfnamefont {R.}~\bibnamefont {Delaney}}, \bibinfo {author} {\bibfnamefont
  {P.}~\bibnamefont {Burns}}, \bibinfo {author} {\bibfnamefont
  {M.}~\bibnamefont {Vissers}}, \bibinfo {author} {\bibfnamefont
  {K.}~\bibnamefont {Lehnert}},\ and\ \bibinfo {author} {\bibfnamefont
  {C.}~\bibnamefont {Regal}},\ }\bibfield  {title} {\bibinfo {title}
  {Optomechanical ground-state cooling in a continuous and efficient
  electro-optic transducer},\ }\href
  {https://doi.org/10.1103/PhysRevX.12.021062} {\bibfield  {journal} {\bibinfo
  {journal} {Physical Review X}\ }\textbf {\bibinfo {volume} {12}},\ \bibinfo
  {pages} {021062} (\bibinfo {year} {2022})}\BibitemShut {NoStop}%
\bibitem [{\citenamefont {Wang}\ \emph {et~al.}(2024)\citenamefont {Wang},
  \citenamefont {Banniard}, \citenamefont {Børkje}, \citenamefont {Massel},
  \citenamefont {Mercier~de Lépinay},\ and\ \citenamefont
  {Sillanpää}}]{Wang2024}%
  \BibitemOpen
  \bibfield  {author} {\bibinfo {author} {\bibfnamefont {C.}~\bibnamefont
  {Wang}}, \bibinfo {author} {\bibfnamefont {L.}~\bibnamefont {Banniard}},
  \bibinfo {author} {\bibfnamefont {K.}~\bibnamefont {Børkje}}, \bibinfo
  {author} {\bibfnamefont {F.}~\bibnamefont {Massel}}, \bibinfo {author}
  {\bibfnamefont {L.}~\bibnamefont {Mercier~de Lépinay}},\ and\ \bibinfo
  {author} {\bibfnamefont {M.~A.}\ \bibnamefont {Sillanpää}},\ }\bibfield
  {title} {\bibinfo {title} {Ground-state cooling of a mechanical oscillator by
  a noisy environment},\ }\href {https://doi.org/10.1038/s41467-024-51645-7}
  {\bibfield  {journal} {\bibinfo  {journal} {Nature Communications}\ }\textbf
  {\bibinfo {volume} {15}},\ \bibinfo {pages} {7395} (\bibinfo {year}
  {2024})}\BibitemShut {NoStop}%
\bibitem [{\citenamefont {Cao}\ \emph {et~al.}(2025)\citenamefont {Cao},
  \citenamefont {Yang}, \citenamefont {Sheng},\ and\ \citenamefont
  {Wu}}]{Cao2025}%
  \BibitemOpen
  \bibfield  {author} {\bibinfo {author} {\bibfnamefont {Y.}~\bibnamefont
  {Cao}}, \bibinfo {author} {\bibfnamefont {C.}~\bibnamefont {Yang}}, \bibinfo
  {author} {\bibfnamefont {J.}~\bibnamefont {Sheng}},\ and\ \bibinfo {author}
  {\bibfnamefont {H.}~\bibnamefont {Wu}},\ }\bibfield  {title} {\bibinfo
  {title} {Optomechanical dark-mode-breaking cooling},\ }\href
  {https://doi.org/10.1103/PhysRevLett.134.043601} {\bibfield  {journal}
  {\bibinfo  {journal} {Physical Review Letters}\ }\textbf {\bibinfo {volume}
  {134}},\ \bibinfo {pages} {043601} (\bibinfo {year} {2025})}\BibitemShut
  {NoStop}%
\bibitem [{\citenamefont {Wise}\ \emph {et~al.}(2024)\citenamefont {Wise},
  \citenamefont {Dutreix},\ and\ \citenamefont {Pistolesi}}]{Wise2024}%
  \BibitemOpen
  \bibfield  {author} {\bibinfo {author} {\bibfnamefont {J.~L.}\ \bibnamefont
  {Wise}}, \bibinfo {author} {\bibfnamefont {C.}~\bibnamefont {Dutreix}},\ and\
  \bibinfo {author} {\bibfnamefont {F.}~\bibnamefont {Pistolesi}},\ }\bibfield
  {title} {\bibinfo {title} {Nonclassical mechanical states in cavity
  optomechanics in the single-photon strong-coupling regime},\ }\href
  {https://doi.org/10.1103/PhysRevA.109.L051501} {\bibfield  {journal}
  {\bibinfo  {journal} {Physical Review A}\ }\textbf {\bibinfo {volume}
  {109}},\ \bibinfo {pages} {l051501} (\bibinfo {year} {2024})}\BibitemShut
  {NoStop}%
\bibitem [{\citenamefont {Bemani}\ \emph {et~al.}(2019)\citenamefont {Bemani},
  \citenamefont {Roknizadeh}, \citenamefont {Motazedifard}, \citenamefont
  {Naderi},\ and\ \citenamefont {Vitali}}]{Bemani2019}%
  \BibitemOpen
  \bibfield  {author} {\bibinfo {author} {\bibfnamefont {F.}~\bibnamefont
  {Bemani}}, \bibinfo {author} {\bibfnamefont {R.}~\bibnamefont {Roknizadeh}},
  \bibinfo {author} {\bibfnamefont {A.}~\bibnamefont {Motazedifard}}, \bibinfo
  {author} {\bibfnamefont {M.~H.}\ \bibnamefont {Naderi}},\ and\ \bibinfo
  {author} {\bibfnamefont {D.}~\bibnamefont {Vitali}},\ }\bibfield  {title}
  {\bibinfo {title} {Quantum correlations in optomechanical crystals},\ }\href
  {https://doi.org/10.1103/PhysRevA.99.063814} {\bibfield  {journal} {\bibinfo
  {journal} {Physical Review A}\ }\textbf {\bibinfo {volume} {99}},\ \bibinfo
  {pages} {063814} (\bibinfo {year} {2019})}\BibitemShut {NoStop}%
\bibitem [{\citenamefont {Massembele}\ \emph {et~al.}(2025)\citenamefont
  {Massembele}, \citenamefont {Djorwé}, \citenamefont {Emale}, \citenamefont
  {Peng}, \citenamefont {Abdel-Aty},\ and\ \citenamefont
  {Nisar}}]{Rostand2025}%
  \BibitemOpen
  \bibfield  {author} {\bibinfo {author} {\bibfnamefont {D.}~\bibnamefont
  {Massembele}}, \bibinfo {author} {\bibfnamefont {P.}~\bibnamefont {Djorwé}},
  \bibinfo {author} {\bibfnamefont {K.}~\bibnamefont {Emale}}, \bibinfo
  {author} {\bibfnamefont {J.-X.}\ \bibnamefont {Peng}}, \bibinfo {author}
  {\bibfnamefont {A.-H.}\ \bibnamefont {Abdel-Aty}},\ and\ \bibinfo {author}
  {\bibfnamefont {K.}~\bibnamefont {Nisar}},\ }\bibfield  {title} {\bibinfo
  {title} {Low threshold quantum correlations via synthetic magnetism in
  brillouin optomechanical system},\ }\href
  {https://doi.org/https://doi.org/10.1016/j.physb.2024.416689} {\bibfield
  {journal} {\bibinfo  {journal} {Physica B: Condensed Matter}\ }\textbf
  {\bibinfo {volume} {697}},\ \bibinfo {pages} {416689} (\bibinfo {year}
  {2025})}\BibitemShut {NoStop}%
\bibitem [{\citenamefont {Emale}\ \emph {et~al.}(2025)\citenamefont {Emale},
  \citenamefont {Peng}, \citenamefont {Djorwé}, \citenamefont {Sarma},
  \citenamefont {Abdourahimi}, \citenamefont {Abdel-Aty}, \citenamefont
  {Nisar},\ and\ \citenamefont {Engo}}]{Emale.2025}%
  \BibitemOpen
  \bibfield  {author} {\bibinfo {author} {\bibfnamefont {K.}~\bibnamefont
  {Emale}}, \bibinfo {author} {\bibfnamefont {J.-X.}\ \bibnamefont {Peng}},
  \bibinfo {author} {\bibfnamefont {P.}~\bibnamefont {Djorwé}}, \bibinfo
  {author} {\bibfnamefont {A.}~\bibnamefont {Sarma}}, \bibinfo {author}
  {\bibnamefont {Abdourahimi}}, \bibinfo {author} {\bibfnamefont {A.-H.}\
  \bibnamefont {Abdel-Aty}}, \bibinfo {author} {\bibfnamefont {K.}~\bibnamefont
  {Nisar}},\ and\ \bibinfo {author} {\bibfnamefont {S.}~\bibnamefont {Engo}},\
  }\bibfield  {title} {\bibinfo {title} {Quantum correlations enhanced in
  hybrid optomechanical system via phase tuning},\ }\href
  {https://doi.org/https://doi.org/10.1016/j.physb.2025.416919} {\bibfield
  {journal} {\bibinfo  {journal} {Physica B: Condensed Matter}\ }\textbf
  {\bibinfo {volume} {701}},\ \bibinfo {pages} {416919} (\bibinfo {year}
  {2025})}\BibitemShut {NoStop}%
\bibitem [{\citenamefont {Slussarenko}\ and\ \citenamefont
  {Pryde}(2019)}]{Slussarenko2019}%
  \BibitemOpen
  \bibfield  {author} {\bibinfo {author} {\bibfnamefont {S.}~\bibnamefont
  {Slussarenko}}\ and\ \bibinfo {author} {\bibfnamefont {G.~J.}\ \bibnamefont
  {Pryde}},\ }\bibfield  {title} {\bibinfo {title} {Photonic quantum
  information processing: A concise review},\ }\href
  {https://doi.org/10.1063/1.5115814} {\bibfield  {journal} {\bibinfo
  {journal} {Applied Physics Reviews}\ }\textbf {\bibinfo {volume} {6}},\
  \bibinfo {pages} {041303} (\bibinfo {year} {2019})}\BibitemShut {NoStop}%
\bibitem [{\citenamefont {Pezzè}\ \emph {et~al.}(2018)\citenamefont {Pezzè},
  \citenamefont {Smerzi}, \citenamefont {Oberthaler}, \citenamefont {Schmied},\
  and\ \citenamefont {Treutlein}}]{Pezz2018}%
  \BibitemOpen
  \bibfield  {author} {\bibinfo {author} {\bibfnamefont {L.}~\bibnamefont
  {Pezzè}}, \bibinfo {author} {\bibfnamefont {A.}~\bibnamefont {Smerzi}},
  \bibinfo {author} {\bibfnamefont {M.~K.}\ \bibnamefont {Oberthaler}},
  \bibinfo {author} {\bibfnamefont {R.}~\bibnamefont {Schmied}},\ and\ \bibinfo
  {author} {\bibfnamefont {P.}~\bibnamefont {Treutlein}},\ }\bibfield  {title}
  {\bibinfo {title} {Quantum metrology with nonclassical states of atomic
  ensembles},\ }\href {https://doi.org/10.1103/RevModPhys.90.035005} {\bibfield
   {journal} {\bibinfo  {journal} {Reviews of Modern Physics}\ }\textbf
  {\bibinfo {volume} {90}},\ \bibinfo {pages} {035005} (\bibinfo {year}
  {2018})}\BibitemShut {NoStop}%
\bibitem [{\citenamefont {Xia}\ \emph {et~al.}(2023)\citenamefont {Xia},
  \citenamefont {Agrawal}, \citenamefont {Pluchar}, \citenamefont {Brady},
  \citenamefont {Liu}, \citenamefont {Zhuang}, \citenamefont {Wilson},\ and\
  \citenamefont {Zhang}}]{Xia2023}%
  \BibitemOpen
  \bibfield  {author} {\bibinfo {author} {\bibfnamefont {Y.}~\bibnamefont
  {Xia}}, \bibinfo {author} {\bibfnamefont {A.~R.}\ \bibnamefont {Agrawal}},
  \bibinfo {author} {\bibfnamefont {C.~M.}\ \bibnamefont {Pluchar}}, \bibinfo
  {author} {\bibfnamefont {A.~J.}\ \bibnamefont {Brady}}, \bibinfo {author}
  {\bibfnamefont {Z.}~\bibnamefont {Liu}}, \bibinfo {author} {\bibfnamefont
  {Q.}~\bibnamefont {Zhuang}}, \bibinfo {author} {\bibfnamefont {D.~J.}\
  \bibnamefont {Wilson}},\ and\ \bibinfo {author} {\bibfnamefont
  {Z.}~\bibnamefont {Zhang}},\ }\bibfield  {title} {\bibinfo {title}
  {Entanglement-enhanced optomechanical sensing},\ }\href
  {https://doi.org/10.1038/s41566-023-01178-0} {\bibfield  {journal} {\bibinfo
  {journal} {Nature Photonics}\ }\textbf {\bibinfo {volume} {17}},\ \bibinfo
  {pages} {470} (\bibinfo {year} {2023})}\BibitemShut {NoStop}%
\bibitem [{\citenamefont {Vitali}\ \emph {et~al.}(2007)\citenamefont {Vitali},
  \citenamefont {Gigan}, \citenamefont {Ferreira}, \citenamefont {Böhm},
  \citenamefont {Tombesi}, \citenamefont {Guerreiro}, \citenamefont {Vedral},
  \citenamefont {Zeilinger},\ and\ \citenamefont {Aspelmeyer}}]{Vitali2007}%
  \BibitemOpen
  \bibfield  {author} {\bibinfo {author} {\bibfnamefont {D.}~\bibnamefont
  {Vitali}}, \bibinfo {author} {\bibfnamefont {S.}~\bibnamefont {Gigan}},
  \bibinfo {author} {\bibfnamefont {A.}~\bibnamefont {Ferreira}}, \bibinfo
  {author} {\bibfnamefont {H.~R.}\ \bibnamefont {Böhm}}, \bibinfo {author}
  {\bibfnamefont {P.}~\bibnamefont {Tombesi}}, \bibinfo {author} {\bibfnamefont
  {A.}~\bibnamefont {Guerreiro}}, \bibinfo {author} {\bibfnamefont
  {V.}~\bibnamefont {Vedral}}, \bibinfo {author} {\bibfnamefont
  {A.}~\bibnamefont {Zeilinger}},\ and\ \bibinfo {author} {\bibfnamefont
  {M.}~\bibnamefont {Aspelmeyer}},\ }\bibfield  {title} {\bibinfo {title}
  {Optomechanical entanglement between a movable mirror and a cavity field},\
  }\href {https://doi.org/10.1103/physrevlett.98.030405} {\bibfield  {journal}
  {\bibinfo  {journal} {Physical Review Letters}\ }\textbf {\bibinfo {volume}
  {98}},\ \bibinfo {pages} {030405} (\bibinfo {year} {2007})}\BibitemShut
  {NoStop}%
\bibitem [{\citenamefont {Paternostro}\ \emph {et~al.}(2007)\citenamefont
  {Paternostro}, \citenamefont {Vitali}, \citenamefont {Gigan}, \citenamefont
  {Kim}, \citenamefont {Brukner}, \citenamefont {Eisert},\ and\ \citenamefont
  {Aspelmeyer}}]{Pater2007}%
  \BibitemOpen
  \bibfield  {author} {\bibinfo {author} {\bibfnamefont {M.}~\bibnamefont
  {Paternostro}}, \bibinfo {author} {\bibfnamefont {D.}~\bibnamefont {Vitali}},
  \bibinfo {author} {\bibfnamefont {S.}~\bibnamefont {Gigan}}, \bibinfo
  {author} {\bibfnamefont {M.~S.}\ \bibnamefont {Kim}}, \bibinfo {author}
  {\bibfnamefont {C.}~\bibnamefont {Brukner}}, \bibinfo {author} {\bibfnamefont
  {J.}~\bibnamefont {Eisert}},\ and\ \bibinfo {author} {\bibfnamefont
  {M.}~\bibnamefont {Aspelmeyer}},\ }\bibfield  {title} {\bibinfo {title}
  {Creating and probing multipartite macroscopic entanglement with light},\
  }\href {https://doi.org/10.1103/physrevlett.99.250401} {\bibfield  {journal}
  {\bibinfo  {journal} {Physical Review Letters}\ }\textbf {\bibinfo {volume}
  {99}},\ \bibinfo {pages} {250401} (\bibinfo {year} {2007})}\BibitemShut
  {NoStop}%
\bibitem [{\citenamefont {Farace}\ and\ \citenamefont
  {Giovannetti}(2012)}]{Farace2012}%
  \BibitemOpen
  \bibfield  {author} {\bibinfo {author} {\bibfnamefont {A.}~\bibnamefont
  {Farace}}\ and\ \bibinfo {author} {\bibfnamefont {V.}~\bibnamefont
  {Giovannetti}},\ }\bibfield  {title} {\bibinfo {title} {Enhancing quantum
  effects via periodic modulations in optomechanical systems},\ }\href
  {https://doi.org/10.1103/PhysRevA.86.013820} {\bibfield  {journal} {\bibinfo
  {journal} {Physical Review A}\ }\textbf {\bibinfo {volume} {86}},\ \bibinfo
  {pages} {013820} (\bibinfo {year} {2012})}\BibitemShut {NoStop}%
\bibitem [{\citenamefont {Mari}\ and\ \citenamefont {Eisert}(2012)}]{Mari2012}%
  \BibitemOpen
  \bibfield  {author} {\bibinfo {author} {\bibfnamefont {A.}~\bibnamefont
  {Mari}}\ and\ \bibinfo {author} {\bibfnamefont {J.}~\bibnamefont {Eisert}},\
  }\bibfield  {title} {\bibinfo {title} {Opto- and electro-mechanical
  entanglement improved by modulation},\ }\href
  {https://doi.org/10.1088/1367-2630/14/7/075014} {\bibfield  {journal}
  {\bibinfo  {journal} {New Journal of Physics}\ }\textbf {\bibinfo {volume}
  {14}},\ \bibinfo {pages} {075014} (\bibinfo {year} {2012})}\BibitemShut
  {NoStop}%
\bibitem [{\citenamefont {Agasti}\ and\ \citenamefont
  {Djorwé}(2024)}]{Agasti2024}%
  \BibitemOpen
  \bibfield  {author} {\bibinfo {author} {\bibfnamefont {S.}~\bibnamefont
  {Agasti}}\ and\ \bibinfo {author} {\bibfnamefont {P.}~\bibnamefont
  {Djorwé}},\ }\bibfield  {title} {\bibinfo {title} {Bistability-assisted
  mechanical squeezing and entanglement},\ }\href
  {https://doi.org/10.1088/1402-4896/ad6eca} {\bibfield  {journal} {\bibinfo
  {journal} {Physica Scripta}\ }\textbf {\bibinfo {volume} {99}},\ \bibinfo
  {pages} {095129} (\bibinfo {year} {2024})}\BibitemShut {NoStop}%
\bibitem [{\citenamefont {Massembele}\ \emph {et~al.}(2024)\citenamefont
  {Massembele}, \citenamefont {Djorwé}, \citenamefont {Sarma}, \citenamefont
  {Abdel-Aty},\ and\ \citenamefont {Engo}}]{Rostand2024}%
  \BibitemOpen
  \bibfield  {author} {\bibinfo {author} {\bibfnamefont {D.~R.~K.}\
  \bibnamefont {Massembele}}, \bibinfo {author} {\bibfnamefont
  {P.}~\bibnamefont {Djorwé}}, \bibinfo {author} {\bibfnamefont {A.~K.}\
  \bibnamefont {Sarma}}, \bibinfo {author} {\bibfnamefont {A.-H.}\ \bibnamefont
  {Abdel-Aty}},\ and\ \bibinfo {author} {\bibfnamefont {S.~G.~N.}\ \bibnamefont
  {Engo}},\ }\bibfield  {title} {\bibinfo {title} {Quantum entanglement
  assisted via duffing nonlinearity},\ }\href
  {https://doi.org/10.1103/PhysRevA.110.043502} {\bibfield  {journal} {\bibinfo
   {journal} {Physical Review A}\ }\textbf {\bibinfo {volume} {110}},\ \bibinfo
  {pages} {043502} (\bibinfo {year} {2024})}\BibitemShut {NoStop}%
\bibitem [{\citenamefont {Lai}\ \emph {et~al.}(2022)\citenamefont {Lai},
  \citenamefont {Chen}, \citenamefont {Qin}, \citenamefont {Miranowicz},\ and\
  \citenamefont {Nori}}]{Lai2022}%
  \BibitemOpen
  \bibfield  {author} {\bibinfo {author} {\bibfnamefont {D.-G.}\ \bibnamefont
  {Lai}}, \bibinfo {author} {\bibfnamefont {Y.-H.}\ \bibnamefont {Chen}},
  \bibinfo {author} {\bibfnamefont {W.}~\bibnamefont {Qin}}, \bibinfo {author}
  {\bibfnamefont {A.}~\bibnamefont {Miranowicz}},\ and\ \bibinfo {author}
  {\bibfnamefont {F.}~\bibnamefont {Nori}},\ }\bibfield  {title} {\bibinfo
  {title} {Tripartite optomechanical entanglement via optical-dark-mode
  control},\ }\href {https://doi.org/10.1103/PhysRevResearch.4.033112}
  {\bibfield  {journal} {\bibinfo  {journal} {Physical Review Research}\
  }\textbf {\bibinfo {volume} {4}},\ \bibinfo {pages} {033112} (\bibinfo {year}
  {2022})}\BibitemShut {NoStop}%
\bibitem [{\citenamefont {Zhai}\ \emph {et~al.}(2023)\citenamefont {Zhai},
  \citenamefont {Du},\ and\ \citenamefont {Guo}}]{Zhai2023}%
  \BibitemOpen
  \bibfield  {author} {\bibinfo {author} {\bibfnamefont {L.-l.}\ \bibnamefont
  {Zhai}}, \bibinfo {author} {\bibfnamefont {H.-J.}\ \bibnamefont {Du}},\ and\
  \bibinfo {author} {\bibfnamefont {J.-L.}\ \bibnamefont {Guo}},\ }\bibfield
  {title} {\bibinfo {title} {Mechanical squeezing and entanglement in coupled
  optomechanical system with modulated optical parametric amplifier},\
  }\bibfield  {journal} {\bibinfo  {journal} {Quantum Information Processing}\
  }\textbf {\bibinfo {volume} {22}},\ \href
  {https://doi.org/10.1007/s11128-023-03965-8} {10.1007/s11128-023-03965-8}
  (\bibinfo {year} {2023})\BibitemShut {NoStop}%
\bibitem [{\citenamefont {Hassan}\ \emph {et~al.}(2015)\citenamefont {Hassan},
  \citenamefont {Hodaei}, \citenamefont {Miri}, \citenamefont {Khajavikhan},\
  and\ \citenamefont {Christodoulides}}]{Hassan2015}%
  \BibitemOpen
  \bibfield  {author} {\bibinfo {author} {\bibfnamefont {A.~U.}\ \bibnamefont
  {Hassan}}, \bibinfo {author} {\bibfnamefont {H.}~\bibnamefont {Hodaei}},
  \bibinfo {author} {\bibfnamefont {M.-A.}\ \bibnamefont {Miri}}, \bibinfo
  {author} {\bibfnamefont {M.}~\bibnamefont {Khajavikhan}},\ and\ \bibinfo
  {author} {\bibfnamefont {D.~N.}\ \bibnamefont {Christodoulides}},\ }\bibfield
   {title} {\bibinfo {title} {Nonlinear reversal of the pt-symmetric phase
  transition in a system of coupled semiconductor microring resonators},\
  }\href {https://doi.org/10.1103/physreva.92.063807} {\bibfield  {journal}
  {\bibinfo  {journal} {Physical Review A}\ }\textbf {\bibinfo {volume} {92}},\
  \bibinfo {pages} {063807} (\bibinfo {year} {2015})}\BibitemShut {NoStop}%
\bibitem [{\citenamefont {Bai}\ \emph {et~al.}(2022)\citenamefont {Bai},
  \citenamefont {Fang}, \citenamefont {Liu}, \citenamefont {Li}, \citenamefont
  {Wan},\ and\ \citenamefont {Xiao}}]{Bai2022}%
  \BibitemOpen
  \bibfield  {author} {\bibinfo {author} {\bibfnamefont {K.}~\bibnamefont
  {Bai}}, \bibinfo {author} {\bibfnamefont {L.}~\bibnamefont {Fang}}, \bibinfo
  {author} {\bibfnamefont {T.-R.}\ \bibnamefont {Liu}}, \bibinfo {author}
  {\bibfnamefont {J.-Z.}\ \bibnamefont {Li}}, \bibinfo {author} {\bibfnamefont
  {D.}~\bibnamefont {Wan}},\ and\ \bibinfo {author} {\bibfnamefont
  {M.}~\bibnamefont {Xiao}},\ }\bibfield  {title} {\bibinfo {title}
  {Nonlinearity-enabled higher-order exceptional singularities with
  ultra-enhanced signal-to-noise ratio},\ }\bibfield  {journal} {\bibinfo
  {journal} {National Science Review}\ }\textbf {\bibinfo {volume} {10}},\
  \href {https://doi.org/10.1093/nsr/nwac259} {10.1093/nsr/nwac259} (\bibinfo
  {year} {2022})\BibitemShut {NoStop}%
\bibitem [{\citenamefont {Gu}\ \emph {et~al.}(2024)\citenamefont {Gu},
  \citenamefont {Qu},\ and\ \citenamefont {Zhang}}]{Gu2024}%
  \BibitemOpen
  \bibfield  {author} {\bibinfo {author} {\bibfnamefont {Q.}~\bibnamefont
  {Gu}}, \bibinfo {author} {\bibfnamefont {C.}~\bibnamefont {Qu}},\ and\
  \bibinfo {author} {\bibfnamefont {Y.}~\bibnamefont {Zhang}},\ }\bibfield
  {title} {\bibinfo {title} {Adjusting exceptional points using saturable
  nonlinearities},\ }\href {https://doi.org/10.1016/j.rinp.2024.107736}
  {\bibfield  {journal} {\bibinfo  {journal} {Results in Physics}\ }\textbf
  {\bibinfo {volume} {61}},\ \bibinfo {pages} {107736} (\bibinfo {year}
  {2024})}\BibitemShut {NoStop}%
\bibitem [{\citenamefont {Lake}\ \emph {et~al.}(2020)\citenamefont {Lake},
  \citenamefont {Mitchell}, \citenamefont {Sanders},\ and\ \citenamefont
  {Barclay}}]{Lake2020}%
  \BibitemOpen
  \bibfield  {author} {\bibinfo {author} {\bibfnamefont {D.~P.}\ \bibnamefont
  {Lake}}, \bibinfo {author} {\bibfnamefont {M.}~\bibnamefont {Mitchell}},
  \bibinfo {author} {\bibfnamefont {B.~C.}\ \bibnamefont {Sanders}},\ and\
  \bibinfo {author} {\bibfnamefont {P.~E.}\ \bibnamefont {Barclay}},\
  }\bibfield  {title} {\bibinfo {title} {Two-colour interferometry and
  switching through optomechanical dark mode excitation},\ }\bibfield
  {journal} {\bibinfo  {journal} {Nature Communications}\ }\textbf {\bibinfo
  {volume} {11}},\ \href {https://doi.org/10.1038/s41467-020-15625-x}
  {10.1038/s41467-020-15625-x} (\bibinfo {year} {2020})\BibitemShut {NoStop}%
\bibitem [{\citenamefont {DeJesus}\ and\ \citenamefont
  {Kaufman}(1987)}]{DeJesus}%
  \BibitemOpen
  \bibfield  {author} {\bibinfo {author} {\bibfnamefont {E.~X.}\ \bibnamefont
  {DeJesus}}\ and\ \bibinfo {author} {\bibfnamefont {C.}~\bibnamefont
  {Kaufman}},\ }\bibfield  {title} {\bibinfo {title} {Routh-hurwitz criterion
  in the examination of eigenvalues of a system of nonlinear ordinary
  differential equations},\ }\href {https://doi.org/10.1103/PhysRevA.35.5288}
  {\bibfield  {journal} {\bibinfo  {journal} {Physical Review A}\ }\textbf
  {\bibinfo {volume} {35}},\ \bibinfo {pages} {5288} (\bibinfo {year}
  {1987})}\BibitemShut {NoStop}%
\bibitem [{\citenamefont {Horodecki}\ \emph {et~al.}(2009)\citenamefont
  {Horodecki}, \citenamefont {Horodecki}, \citenamefont {Horodecki},\ and\
  \citenamefont {Horodecki}}]{Horodecki2009}%
  \BibitemOpen
  \bibfield  {author} {\bibinfo {author} {\bibfnamefont {R.}~\bibnamefont
  {Horodecki}}, \bibinfo {author} {\bibfnamefont {P.}~\bibnamefont
  {Horodecki}}, \bibinfo {author} {\bibfnamefont {M.}~\bibnamefont
  {Horodecki}},\ and\ \bibinfo {author} {\bibfnamefont {K.}~\bibnamefont
  {Horodecki}},\ }\bibfield  {title} {\bibinfo {title} {Quantum entanglement},\
  }\href {https://doi.org/10.1103/RevModPhys.81.865} {\bibfield  {journal}
  {\bibinfo  {journal} {Reviews of Modern Physics}\ }\textbf {\bibinfo {volume}
  {81}},\ \bibinfo {pages} {865} (\bibinfo {year} {2009})}\BibitemShut
  {NoStop}%
\bibitem [{\citenamefont {Uola}\ \emph {et~al.}(2020)\citenamefont {Uola},
  \citenamefont {Costa}, \citenamefont {Nguyen},\ and\ \citenamefont
  {Gühne}}]{Uola2020}%
  \BibitemOpen
  \bibfield  {author} {\bibinfo {author} {\bibfnamefont {R.}~\bibnamefont
  {Uola}}, \bibinfo {author} {\bibfnamefont {A.~C.}\ \bibnamefont {Costa}},
  \bibinfo {author} {\bibfnamefont {H.~C.}\ \bibnamefont {Nguyen}},\ and\
  \bibinfo {author} {\bibfnamefont {O.}~\bibnamefont {Gühne}},\ }\bibfield
  {title} {\bibinfo {title} {Quantum steering},\ }\href
  {https://doi.org/10.1103/RevModPhys.92.015001} {\bibfield  {journal}
  {\bibinfo  {journal} {Reviews of Modern Physics}\ }\textbf {\bibinfo {volume}
  {92}},\ \bibinfo {pages} {015001} (\bibinfo {year} {2020})}\BibitemShut
  {NoStop}%
\bibitem [{\citenamefont {Xiang}\ \emph {et~al.}(2022)\citenamefont {Xiang},
  \citenamefont {Cheng}, \citenamefont {Gong}, \citenamefont {Ficek},\ and\
  \citenamefont {He}}]{Xiang2022}%
  \BibitemOpen
  \bibfield  {author} {\bibinfo {author} {\bibfnamefont {Y.}~\bibnamefont
  {Xiang}}, \bibinfo {author} {\bibfnamefont {S.}~\bibnamefont {Cheng}},
  \bibinfo {author} {\bibfnamefont {Q.}~\bibnamefont {Gong}}, \bibinfo {author}
  {\bibfnamefont {Z.}~\bibnamefont {Ficek}},\ and\ \bibinfo {author}
  {\bibfnamefont {Q.}~\bibnamefont {He}},\ }\bibfield  {title} {\bibinfo
  {title} {Quantum steering: Practical challenges and future directions},\
  }\href {https://doi.org/10.1103/PRXQuantum.3.030102} {\bibfield  {journal}
  {\bibinfo  {journal} {PRX Quantum}\ }\textbf {\bibinfo {volume} {3}},\
  \bibinfo {pages} {030102} (\bibinfo {year} {2022})}\BibitemShut {NoStop}%
\bibitem [{\citenamefont {Fang}\ \emph {et~al.}(2025)\citenamefont {Fang},
  \citenamefont {Bai}, \citenamefont {Guo}, \citenamefont {Liu}, \citenamefont
  {Li},\ and\ \citenamefont {Xiao}}]{Fang2025}%
  \BibitemOpen
  \bibfield  {author} {\bibinfo {author} {\bibfnamefont {L.}~\bibnamefont
  {Fang}}, \bibinfo {author} {\bibfnamefont {K.}~\bibnamefont {Bai}}, \bibinfo
  {author} {\bibfnamefont {C.}~\bibnamefont {Guo}}, \bibinfo {author}
  {\bibfnamefont {T.-R.}\ \bibnamefont {Liu}}, \bibinfo {author} {\bibfnamefont
  {J.-Z.}\ \bibnamefont {Li}},\ and\ \bibinfo {author} {\bibfnamefont
  {M.}~\bibnamefont {Xiao}},\ }\bibfield  {title} {\bibinfo {title}
  {Exceptional features in nonlinear hermitian systems},\ }\href
  {https://doi.org/10.1103/PhysRevB.111.L161102} {\bibfield  {journal}
  {\bibinfo  {journal} {Physical Review B}\ }\textbf {\bibinfo {volume}
  {111}},\ \bibinfo {pages} {l161102} (\bibinfo {year} {2025})}\BibitemShut
  {NoStop}%
\end{thebibliography}%

\end{document}